\documentclass[aps,prm,twocolumn,superscriptaddress,showpacs,english,nofootinbib]{revtex4-1}
\usepackage{caption}
\usepackage{float}
\usepackage{subcaption}
\usepackage[T1]{fontenc}
\usepackage[utf8]{inputenc}
\usepackage{babel}
\usepackage{amsmath}
\usepackage{amssymb}
\usepackage{wasysym}
\usepackage{graphicx}
\usepackage{multirow}
\usepackage{gensymb}
\usepackage[dvipsnames]{xcolor}
\usepackage{pifont}
\usepackage{microtype}
\usepackage{xcolor}
\usepackage{soul}
\usepackage[version=3]{mhchem} 
\usepackage{chemformula}[2014/04/08] 
\usepackage{booktabs}    
\usepackage{dcolumn}     
\usepackage{bm}  

\usepackage[linktocpage=true,
  colorlinks=true, 
  pdfborder={0 0 0},
  linkcolor=blue,
  citecolor=red,
  filecolor=yellow,
  urlcolor=blue,
  bookmarks,
  pdfauthor={},
]{hyperref}

\begin{document}

\title{Maximum volume simplex method for automatic selection and classification of atomic environments and  environment descriptor compression} 
\author{Behnam Parsaeifard}
\author{Daniele Tomerini}
\author{Deb Sankar De}
\author{Stefan Goedecker}
\affiliation{Department\ of\ Physics,\ Universitaet\ Basel,\ Klingelbergstrasse\ 82,\ 4056\ Basel,\ Switzerland}
\affiliation{National Center for Computational Design and Discovery of Novel Materials (MARVEL), Switzerland} 
\date{\today}

\begin{abstract}

Fingerprint distances, which measure the similarity of atomic environments, are commonly calculated from atomic environment fingerprint vectors. In this work we present the simplex method which can 
perform the inverse operation, i.e. calculating fingerprint vectors from fingerprint distances. 
The fingerprint vectors found in this way point to the corners of a simplex.
For a large data set of fingerprints, we can find a particular largest volume simplex, 
whose dimension gives the effective dimension of the fingerprint vector space. 
We show that the corners of this simplex correspond to landmark environments that can 
by used in a fully automatic way to analyse structures. In this way we can for instance 
detect atoms in grain boundaries or on edges of carbon flakes without any human input about 
the expected environment. By projecting fingerprints on the largest volume simplex we can also 
obtain fingerprint vectors that are considerably shorter than the original ones but whose 
information content is not significantly reduced.

\end{abstract} 

\maketitle


\section{Introduction}
Materials science has become to a large extent a data driven science.
Several data banks exist that contains not only structural data, but calculated properties as well; many exceed the hundreds of thousands structural properties in number, with their number growing dramatically~\cite{ceder,wolverton,curtarolo,Talirz2020}.
Molecular dynamics simulations typically also generate very large data sets.
Such large data sets can not any more be inspected by eye and tools for classifying the 
structures in an automatic way are needed. 
Atomic environments can be described in a quantitative fashion by descriptors called "atomic environment fingerprints"~\cite{bartok2013representing,behler2011atom,faber2018alchemical,christensen2019fchl,mallat}, that can also provide a description for entire crystalline structures~\cite{zhu2016fingerprint}.
Atomic environment fingerprints are 
also used as inputs for supervised
machine learning schemes~\cite{behler2015constructing,rupp2012fast,bartok2010gaussian} of potential energy surfaces. 
For such a use
it is desirable that the fingerprint is able to detect any difference in the environment~\cite{parsaeifard2020assessment} 
while keeping
the fingerprint vector as short as possible. 

One of our goals  
will be the detection of grain boundaries, that are the disordered regions between one or two  ordered phases.
Grain boundaries have an important influence on physical properties of the system including strength, conductivity, ductility, and crack resistance to name but a few ~\cite{hansen2004hall,chiba1994relation,fang2011revealing,shimada2002optimization,lu2004ultrahigh,meyers2006mechanical}. 

Several methods have been proposed in the literature to distinguish between certain reference crystalline 
structures and disordered and mainly liquid structures in melting and nucleation 
simulations such as Steinhardt parameters~\cite{steinhardt} and 
common neighbour analysis (CNA) ~\cite{faken1994systematic}.
These methods have also been used to study dislocations, local ordering and grain boundaries ~\cite{schiotz2003maximum,yamakov2003deformation,brandl2011dislocation,jonsson1988icosahedral,bailey2004simulation}. 
One of the disadvantages of these methods is that they are based on a sharp cutoff, and they end up lacking smoothness  
with respect to particle displacements occurring in MD or during relaxations. As its name suggests, in the adaptive common neighbour analysis~\cite{stukowski2012structure} the cutoff is adapted to the environment of each atom. Although 
more robust compared to CNA, it remains sensitive to thermal vibrations. Different predefined crystalline structures can be distinguished by polyhedral template matching ~\cite{larsen2016robust}. SOAP \cite{bartok2013representing} fingerprints coupled to machine learning methods were recently also used to predict properties of grain boundaries \cite{rosenbrock2017discovering}.
Based on a formula to calculate the entropy for a system interacting only 
via pairwise forces, an atomic entropy can be obtained which allows to distinguish 
between liquid, FCC, BCC and HCP crystalline phases~\cite{piaggi2017entropy}. 

The common characteristic of all existing methods is
that they require some human input about the relevant structures that are expected to be encountered in the simulation. This is in contrast to our method which selects all the relevant structures fully automatically based on a large pool of structures. 
The method is also applicable without any adjustments to any molecular system.


\section{The largest volume simplex method} \label{simplex_method}
\subsection{fingerprints and fingerprint distances}
In this section we provide a short review of the overlap matrix fingerprint method, that we use to describe the local atomic environment. A complete description can be find in the original paper detailing the method~\cite{zhu2016fingerprint}.


In order to calculate the overlap matrix (OM) fingerprint for an atom $k$ in a structure, we take into account the relative position of all the neighbours of that atom within a cutoff sphere (centered on atom $k$) of radius $R_c$. Neighbours include all the relevant periodic images of an atom when dealing with an atom at the edge of a repeating unit for a periodic system.
To each one of the atoms is associated a minimal set of normalized atom-centered Gaussians $G_\nu(\bm{r}-\bm{R}_i)$, centered on the atom itself. The width of each Gaussian is given by the covalent radius of the atom on which it is centered.
For carbon with its strong directional bonding we have used a set of s and p-type orbitals ($\nu=s,p_x,p_y,p_z$)
and denote the resulting fingerprint by OM[sp], for aluminum with its metallic bonding 
we have used only $\nu=s$ and denote the fingerprint by OM[s]. 
We then calculate the overlap between Gaussian functions in the sphere.
\begin{equation}
    S^k_{i,\nu,j,\mu}=\int G_\nu(\bm{r}-\bm{R}_i)G_\mu(\bm{r}-\bm{R}_j)d\bm{r}
\end{equation}
Next, the overlap matrix $S^k_{i,\nu,j,\mu}$ is multiplied by the amplitude functions $f_c(|\bm{R}_k-\bm{R}_i|)$ and $f_c(|\bm{R}_k-\bm{R}_j|)$ to obtain a modified overlap 
matrix $\tilde S$
\begin{equation}
    \tilde{S}^k_{i,\nu,j,\mu}=f_c(|\bm{R}_k-\bm{R}_i)S^k_{i,\nu,j,\mu}f_c(|\bm{R}_k-\bm{R}_j|)
\end{equation}
$f_c(r)=(1-\frac{r^2}{4 w^2})^2$ is a cutoff function that vanishes at and beyond $r=2w=R_c$ 
with two continuous derivatives.
$w$ gives the length scale over which $f_c(r)$ drops to zero and we typically choose it so that about 50 atoms are contained within the cutoff radius $R_c$. 
The matrix whose columns are denoted by the composite index $i,\nu$ and whose rows are 
given by the composite index $j,\mu$ is then diagonalized to obtain the eigenvalues.
Finally, the vector $\bm {V}^k$ containing all the eigenvalues of the matrix $\tilde{S}^k_{i,\nu,j,\mu}$ is the fingerprint of atom $k$. 
It has a length $L=4 N_{sphere}$ for OM[sp]
and $L=N_{sphere}$ for OM[s] where $N_{sphere}$ is the number of atoms in the sphere around the central atom. 

By construction the fingerprint is robust against displacements of the atoms across the boundary of the sphere radius, and therefore it is possible to calculate derivatives of the fingerprints with respect to infinitesimal structural change around the atom $k$.
The fingerprint vectors $\bm{V}^k$ characterize the atomic environments around atom $k$ and the 
fingerprint distance $d_{i,j}$ is a measure of the dissimilarity between two environments $i$ and $j$. The fingerprint distance is obtained from the euclidean norm of the difference vector throughout this study:  
\begin{equation} \label{eq:fingerprint_distance}
    d_{i,j}=|\bm{V}^i-\bm{V}^j|
\end{equation}

\subsection{Obtaining fingerprint vectors from fingerprint distances}
The above formula~\ref{eq:fingerprint_distance} gives a trivial recipe to obtain fingerprint distances $d_{i,j}$ from a set of 
points represented by the fingerprint vectors in a space of dimension $L$. In the following we will derive the formulas for 
the inverse operation. Given a set of pairwise fingerprint distances $d_{i,j}$ we want to construct a set of points 
$\bm{x}^i $ that will satisfy these constraints.
The solution of this problem is not unique. 
The solution can however be made unique by requiring that the first
point be the origin, $\bm x^0=0$, and that for each consecutive point the number of 
nonzero components increases by one. Hence the  points $\bm x^i$
have the following structure:
\begin{equation} \label{eq_matrix}
(\bm x^1, \bm x^2, \dots, \bm x^N)=
\begin{pmatrix}
x_{1,1} & x_{1,2} & \hdots & x_{1,N}\\
0      & x_{2,2} & \hdots & x_{2,N}\\
\vdots & \vdots & \ddots & \vdots\\
0 & 0 & 0 & x_{N,N}
\end{pmatrix}  
\end{equation}
So after placing the first point at the origin, the next point lies on the positive x-axis at the right distance, the following on the xy plane (y>0), and so on. 
The components of the set of points $\bm x^i$'s can be obtained recursively from  simple relations between the distances among the vectors $\bm V^i$'s.

The distance between $\bm x^N$ and the origin, $\bm x^0$, is simply given by the norm of the vector:
\begin{equation}
    d_{0,N}^2 = \sum_{i=1}^{N}{x_{i,N}^2}
\end{equation}
For $M < N$, the difference between column $N$ and $M$ is related to the distance between points $\bm x^N$ and $\bm x^M$ as
\begin{equation}
    d_{M,N}^2 = \sum_{i=1}^{M}{(x_{i,N}-x_{i,M})^2} + \sum_{j=M+1}^{N}{x_{j,N}^2}
\end{equation}
By taking the difference between $d^2_{M,N}$ and $d^2_{0,N}$ we obtain a simplified set of equations:
\begin{multline}\label{eq4}
     d_{M,N}^2-d_{0,N}^2 = \sum_{i=1}^{M}{(x_{i,N}-x_{i,M})^2-x_{i,N}^2} \\
     =\sum_{i=1}^{M}{-x_{i,M}(2x_{i,N}-x_{i,M})}
\end{multline}

In Eq.~\ref{eq4}, the unknowns $x_{i,N}$ depends only on other column elements $x_{j,M}$ with $M<N$.
\begin{equation} \label{eq:1}
    x_{1,1}=d_{0,1}
\end{equation}
\begin{equation} \label{eq:2}
    x_{1,2}=\frac{d^2_{0,1}+d^2_{0,2}-d^2_{1,2}}{2x_{1,1}}
\end{equation}
\begin{equation} \label{eq:3}
    x_{2,2}=\sqrt{d^2_{0,2}-x^2_{1,2}}
\end{equation}
We can write for $M<N$ in general:
\begin{equation} \label{eq:4}
    x_{M,N}= \frac{d^2_{0,M}+d^2_{0,N}-d^2_{M,N}-2\sum_{i=1}^{M-1}{x_{i,M}x_{i,N}} }{2x_{M,M}}
\end{equation}
and for $M=N$ we have:
\begin{equation}
    x_{N,N}=\sqrt{d^2_{0,N}-\sum_{i=1}^{N-1}{x^2_{i,N}}} \label{eq:lastx}
\end{equation}
The geometrical body having as corners the above calculated points is 
a $N$-dimensional simplex with volume $x_{1,1} x_{2,2} \dots  x_{N,N}/N!$. 
The above construction can be done for any set of $\frac{N_{env}(N_{env}-1)}{2}$ distances 
as long as the original $\bm V^i$'s 
giving rise to the distances via Eq.~\ref{eq:fingerprint_distance} are linearly independent. Since the number of environments $N_{env}$ is typically much larger than the length $L$ of the fingerprint vectors, at most $L$ points (including in the count the origin) can be obtained. If  the number of linearly independent 
fingerprint vectors is less than $L$, $x_{i,i}$ will become zero for some $i < L$ and it is thus 
not possible to increase the dimension of the simplex.
In the context of our fingerprints, it turns out that the $x_{i,i}$ typically are not exactly zero but  become 
very small which means that all the fingerprint vectors are essentially contained in a sub volume whose 
dimension is smaller than $L$. The component that is orthogonal to this subspace is then very small 
and can be neglected. This is the basic property which will be exploited for the fingerprint 
compression later in the paper.

\subsection{Construction of the largest volume simplex} \label{largestvolumesimplexconstruction}
Now, we will describe how we can use the construction outlined above to
obtain the largest volume simplex which we will simply denote by largest simplex (LS).
We do this since we are interested to find the effective dimension $l$ of the space spanned by the fingerprints which gives the number of the highly distinctive landmark environments 
together with these environments.
We start by identifying the two environments characterized by the largest distance. 
This defines the origin $\bm x^0$ and the first point along the x-axis, i.e. $\bm x^1$, and in this 
way the first two corners of the simplex, which is at this stage just a line.
To enlarge in the next step the dimension of the simplex by one we search for the environment that 
will give the largest area triangle if the point $\bm x^2$, corresponding to this environment, is used as 
the third corner. We then increase the dimension of the simplex step by step and we choose the new corners in each step in such a way that the volume of the new simplex will be maximal. The procedure is stopped if in a certain step $l$ the volume collapses to a very small value because additional fingerprint vectors are 
quasi linearly dependent on the previous ones. In this way an effective dimension $l$ of the entire fingerprint space can be determined. 
Once this maximum volume simplex is constructed we can express other fingerprint vectors in the basis of the vectors $\bm x^i$ spanning the LS simplex. To get the expansion coefficients, we just perform the same steps of Eqs.~\ref{eq:1} to ~\ref{eq:lastx}.
that would be needed to add a corner to the simplex. However in this case we know already that the 
$x_{l+1,l+1}$ from Eq.~\ref{eq:lastx} will be negligible because we stopped the maximum volume simplex construction exactly for the reason that we could not find any point that gave a large  $x_{l+1,l+1}$.

\section{Applications}
 
In this section, we show some applications of the LS. 
In section \ref{sub_C60} we apply the methodology to the study of a variety of C$_{60}$ molecules, to identify the most distinct environments and group the most similar ones.
In section \ref{sub_Al} we use the method to find the grain boundaries in a Al nanocrystalline material.
In~\ref{sec:discussion} we exploit the LS to reduce the dimensions of the fingerprint and compare its performance with CUR decomposition method \cite{mahoney2009cur}.

 
\subsection{$C_{60}$ clusters} \label{sub_C60}
\begin{figure*}[p]
     \centering
     \begin{subfigure}[b]{0.2\textwidth}
         \centering
         \includegraphics[width=\textwidth]{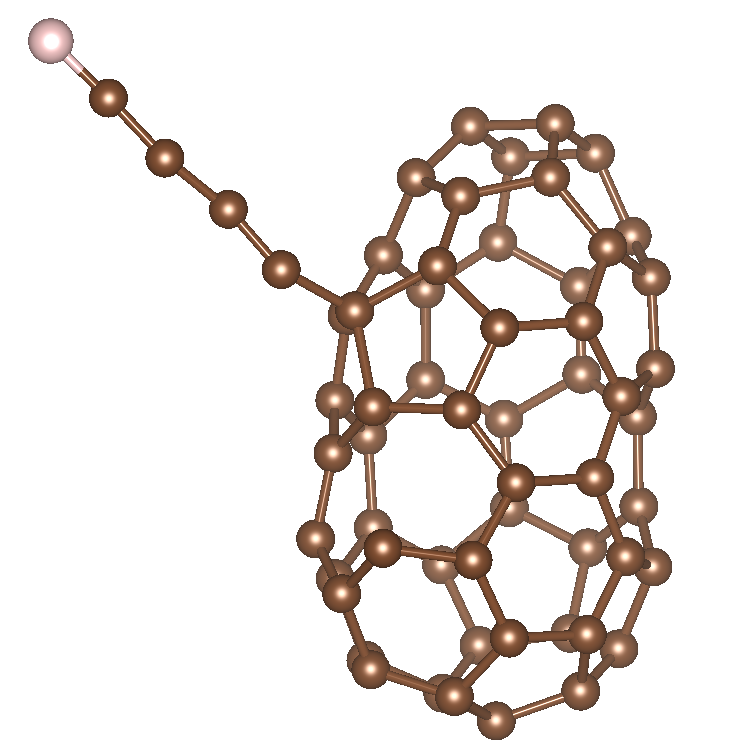}
         \caption{}
         \label{fig:c60corner_000}
     \end{subfigure}
     \begin{subfigure}[b]{0.2\textwidth}
         \centering
         \includegraphics[width=\textwidth]{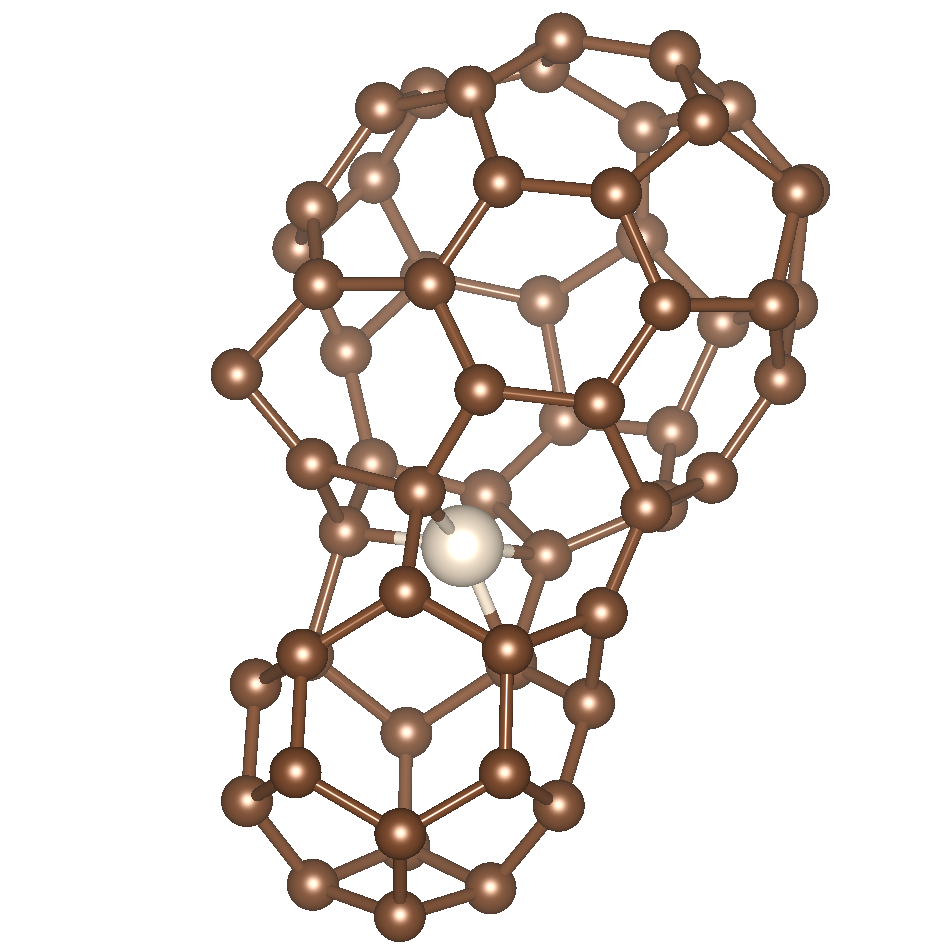}
         \caption{}
         \label{fig:c60corner_001}
     \end{subfigure}
     \begin{subfigure}[b]{0.2\textwidth}
         \centering
         \includegraphics[width=\textwidth]{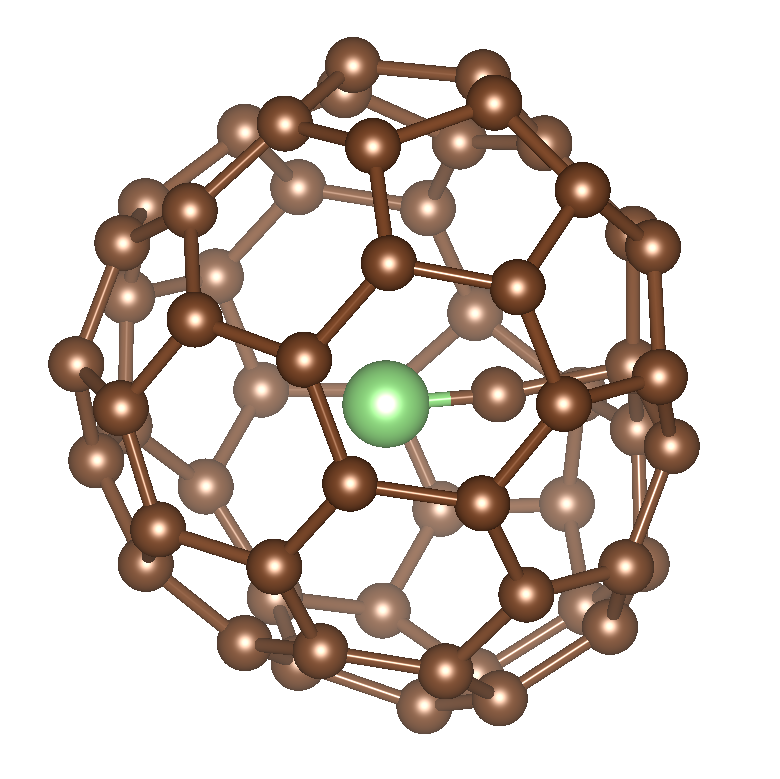}
         \caption{}
         \label{fig:c60corner_002}
     \end{subfigure}
     \begin{subfigure}[b]{0.2\textwidth}
         \centering
         \includegraphics[width=\textwidth]{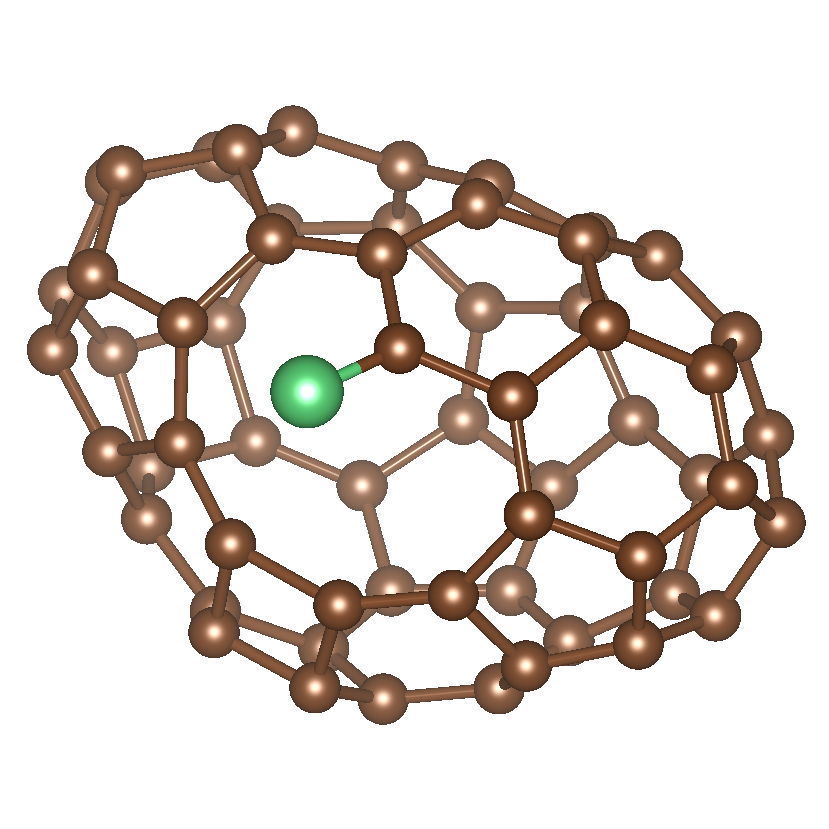}
         \caption{}
         \label{fig:c60corner_003}
     \end{subfigure}
     \begin{subfigure}[b]{0.2\textwidth}
         \centering
         \includegraphics[width=\textwidth]{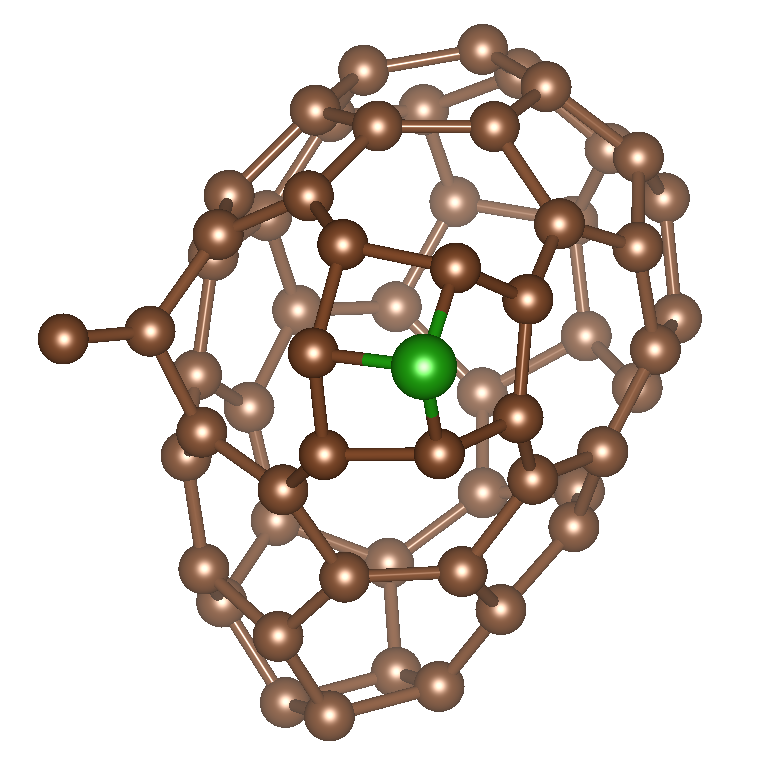}
         \caption{}
         \label{fig:c60corner_004}
     \end{subfigure}
     \begin{subfigure}[b]{0.2\textwidth}
         \centering
         \includegraphics[width=\textwidth]{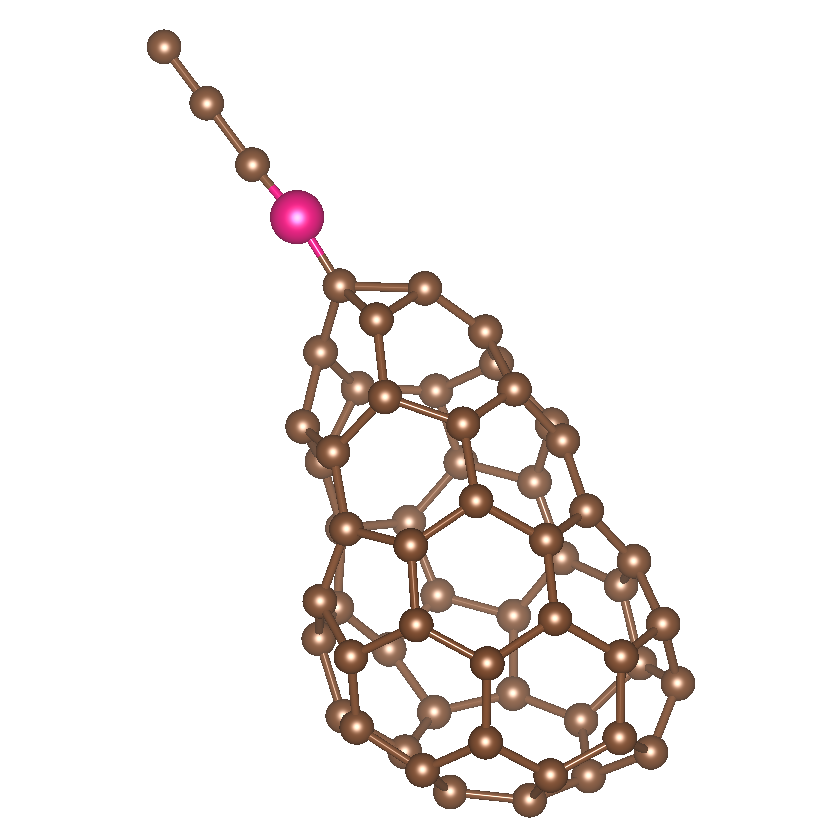}
         \caption{}
         \label{fig:c60corner_005}
     \end{subfigure}
      \begin{subfigure}[b]{0.2\textwidth}
         \centering
         \includegraphics[width=\textwidth]{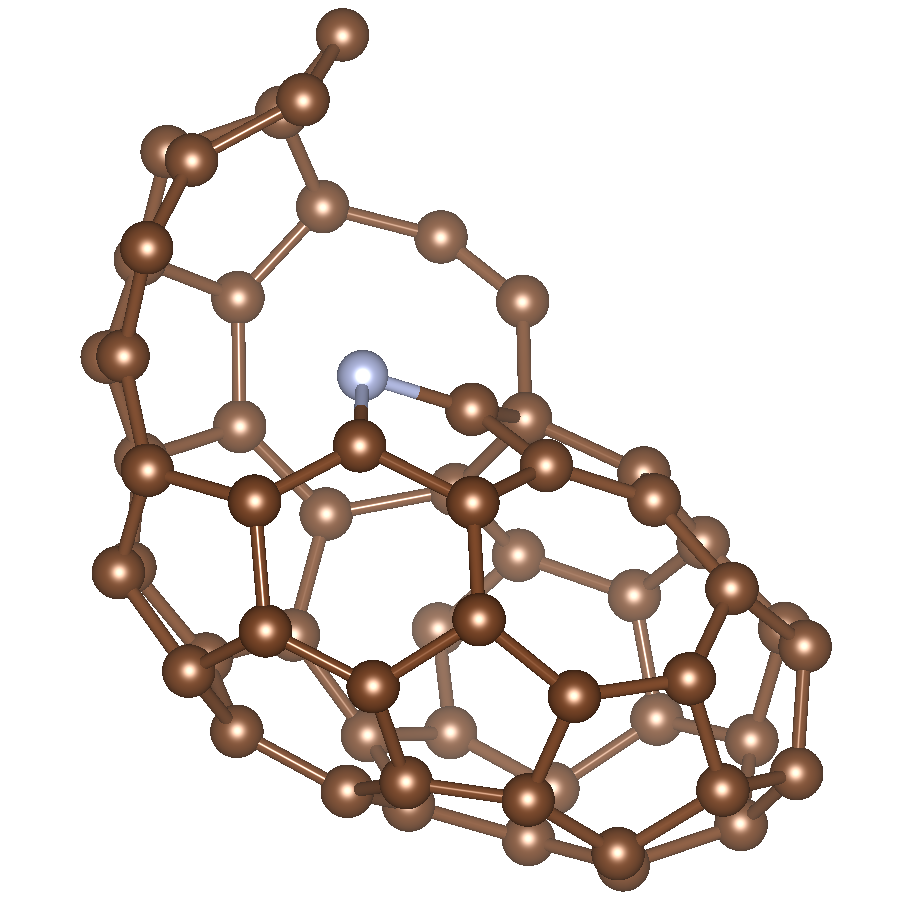}
         \caption{}
         \label{fig:c60corner_006}
     \end{subfigure}
      \begin{subfigure}[b]{0.2\textwidth}
         \centering
         \includegraphics[width=\textwidth]{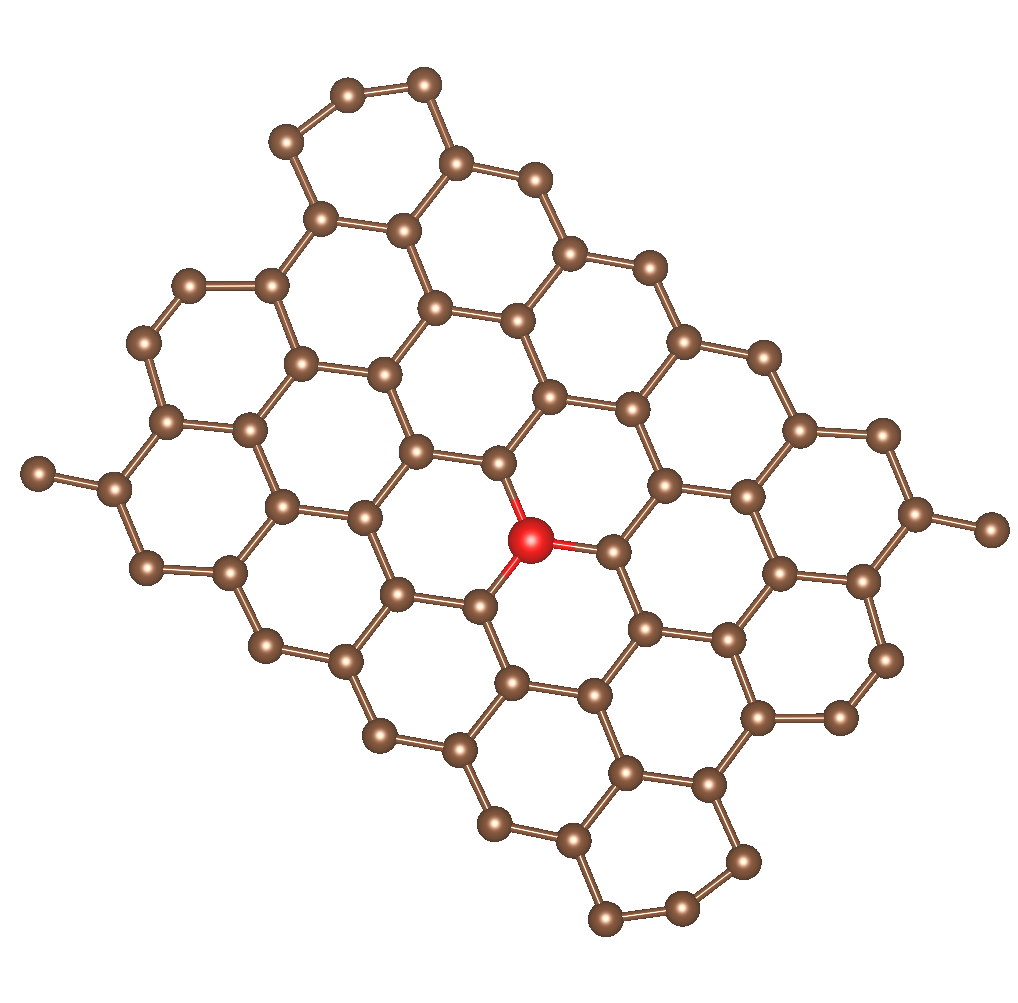}
         \caption{}
         \label{fig:c60corner_007}
     \end{subfigure}
      \begin{subfigure}[b]{0.2\textwidth}
         \centering
         \includegraphics[width=\textwidth]{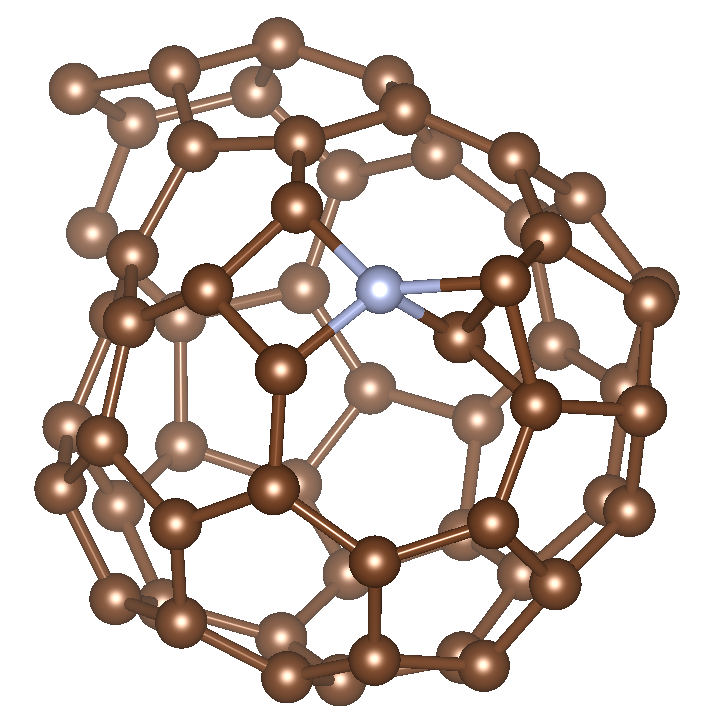}
         \caption{}
         \label{fig:c60corner_008}
     \end{subfigure}
      \begin{subfigure}[b]{0.2\textwidth}
         \centering
         \includegraphics[width=\textwidth]{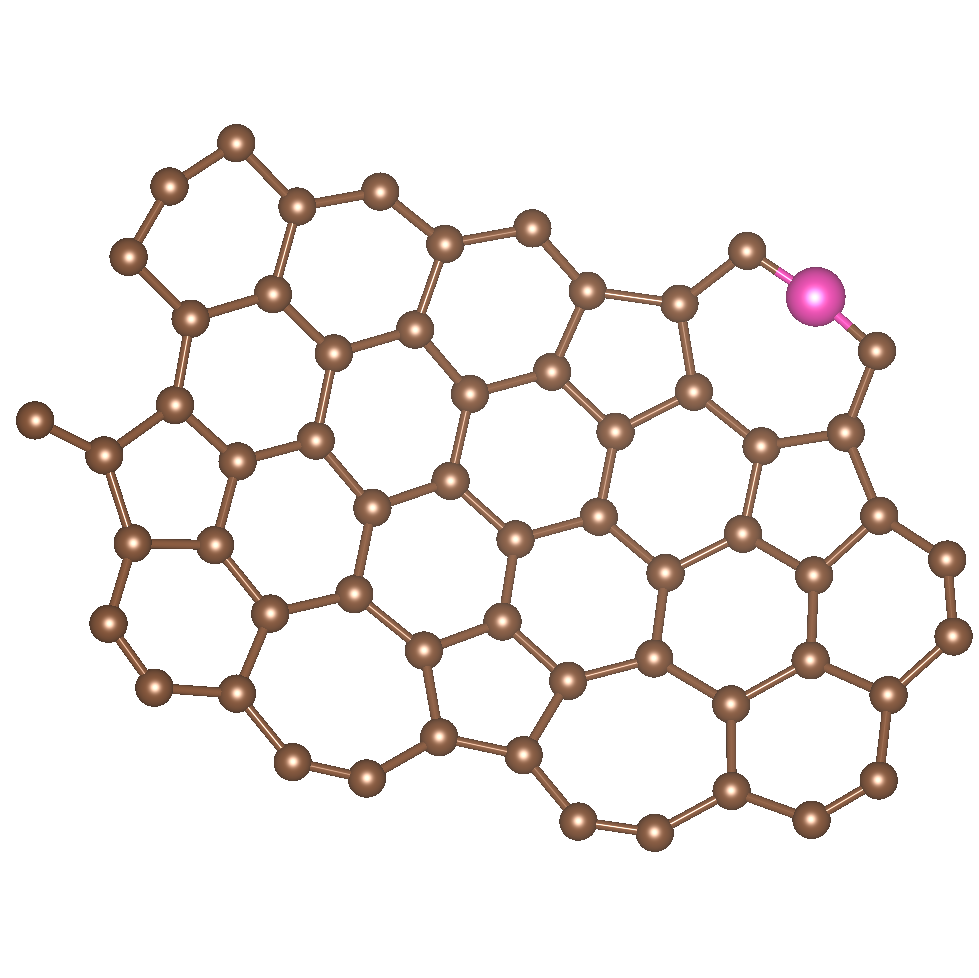}
         \caption{}
         \label{fig:c60corner_009}
     \end{subfigure}
      \begin{subfigure}[b]{0.2\textwidth}
         \centering
         \includegraphics[width=\textwidth]{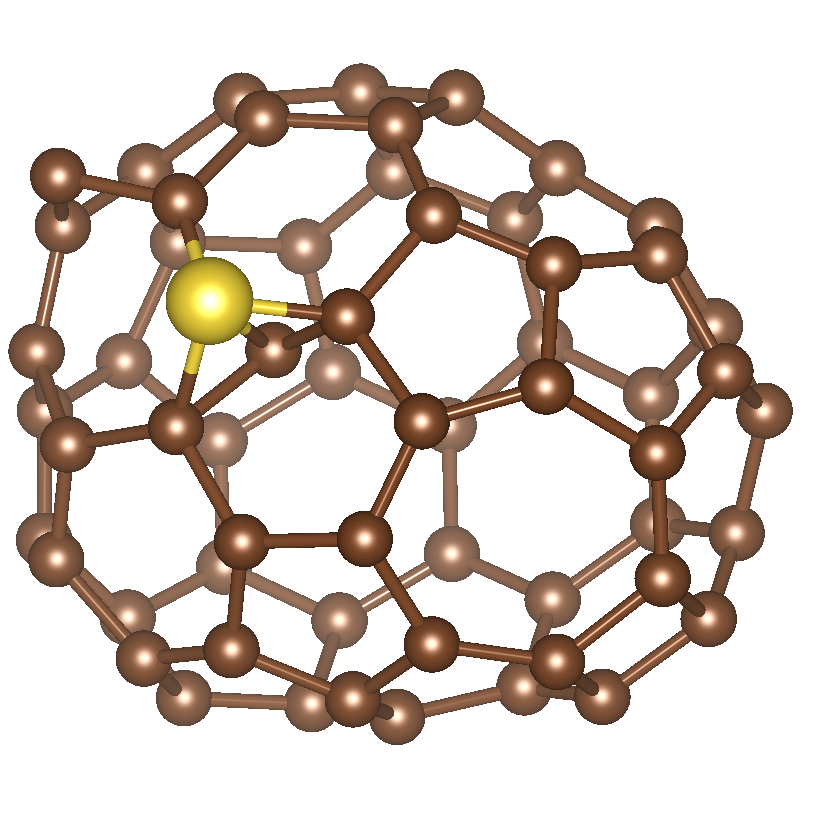}
         \caption{}
         \label{fig:c60corner_010}
     \end{subfigure}
      \begin{subfigure}[b]{0.2\textwidth}
         \centering
         \includegraphics[width=\textwidth]{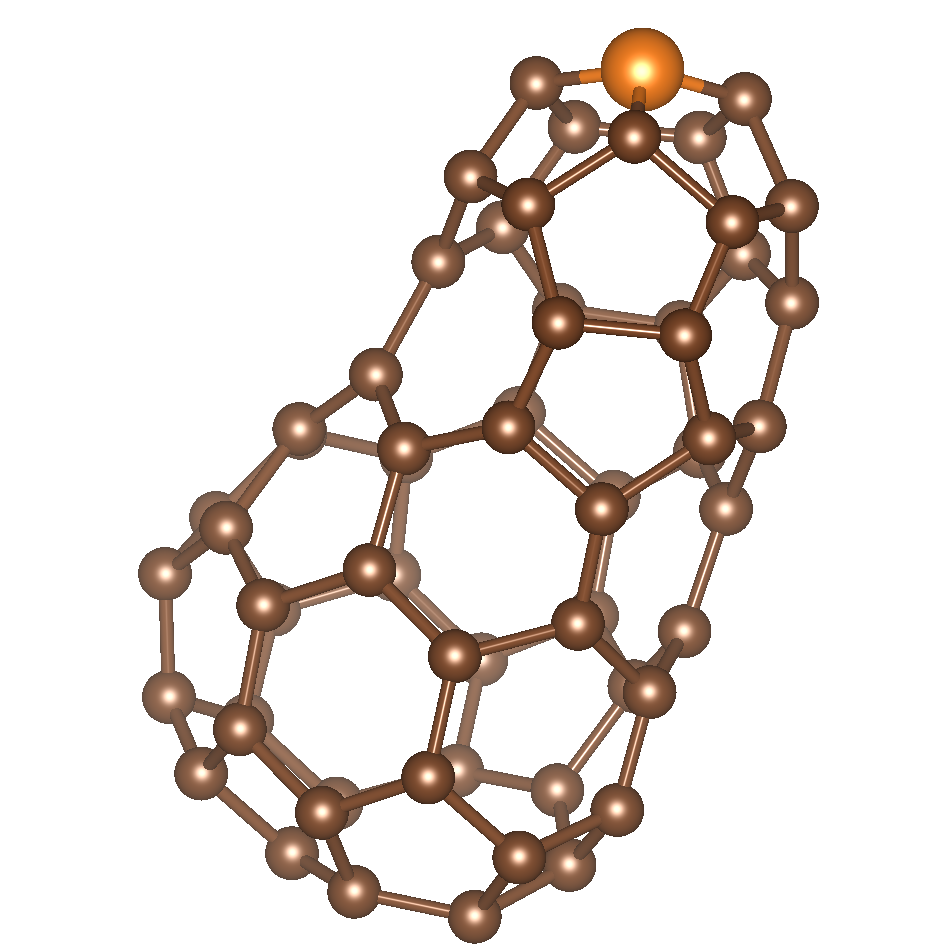}
         \caption{}
         \label{fig:c60corner_011}
     \end{subfigure}
      \begin{subfigure}[b]{0.2\textwidth}
         \centering
         \includegraphics[width=\textwidth]{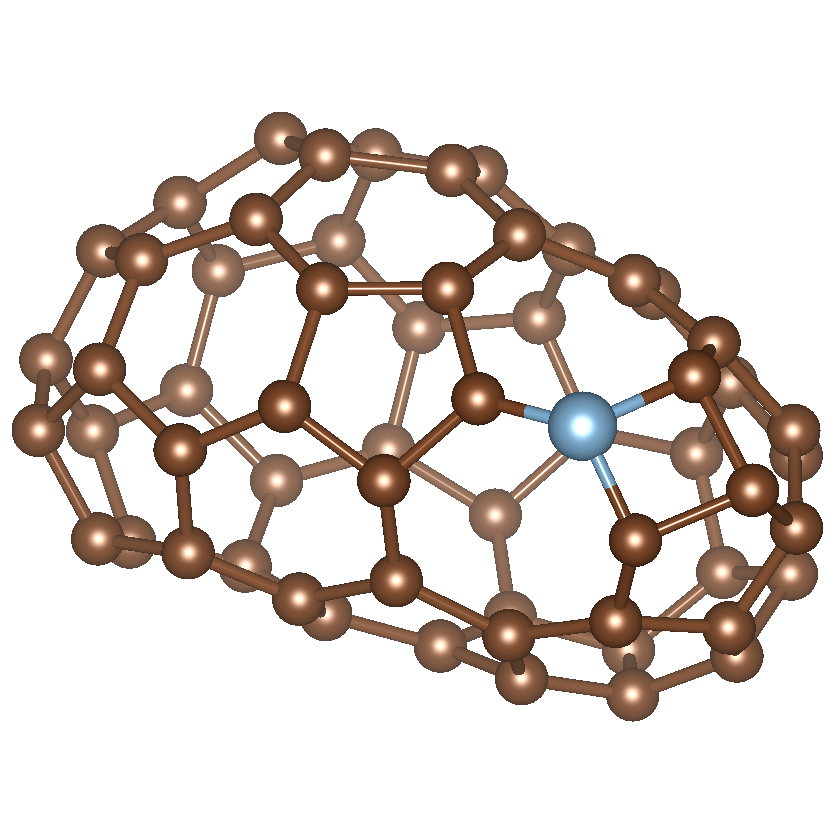}
         \caption{}
         \label{fig:c60corner_012}
     \end{subfigure}
     \begin{subfigure}[b]{0.2\textwidth}
         \centering
         \includegraphics[width=\textwidth]{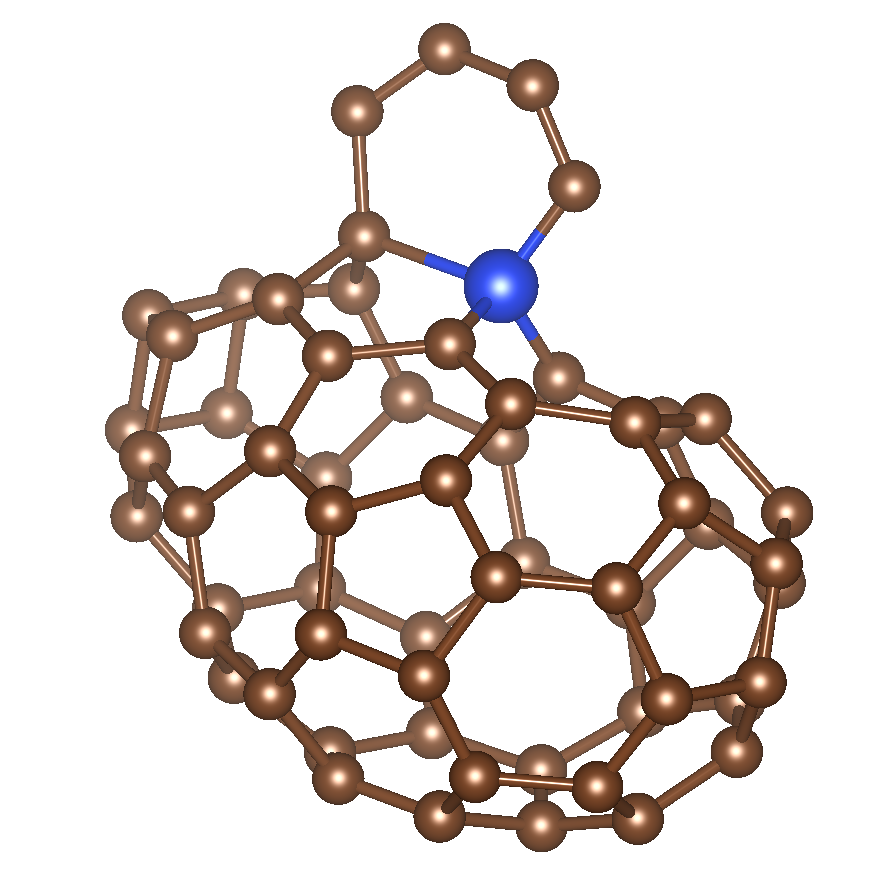}
         \caption{}
         \label{fig:c60corner_013}
     \end{subfigure}
     \begin{subfigure}[b]{0.2\textwidth}
         \centering
         \includegraphics[width=\textwidth]{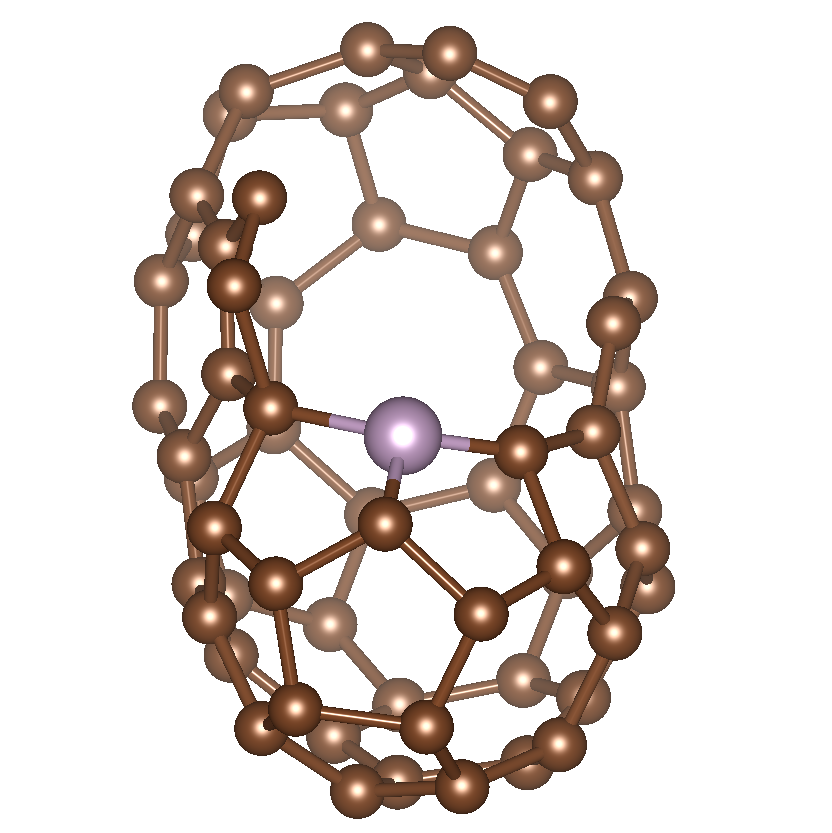}
         \caption{}
         \label{fig:c60corner_014}
     \end{subfigure}
     \begin{subfigure}[b]{0.2\textwidth}
         \centering
         \includegraphics[width=\textwidth]{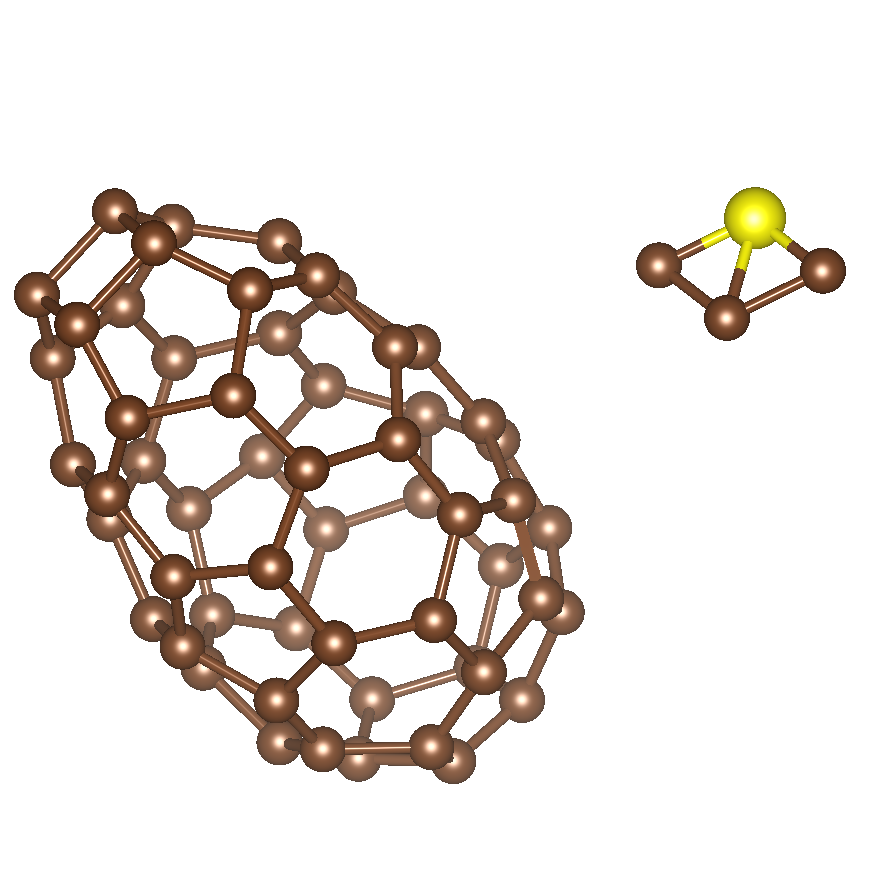}
         \caption{}
         \label{fig:c60corner_015}
     \end{subfigure}
     \begin{subfigure}[b]{0.2\textwidth}
         \centering
         \includegraphics[width=\textwidth]{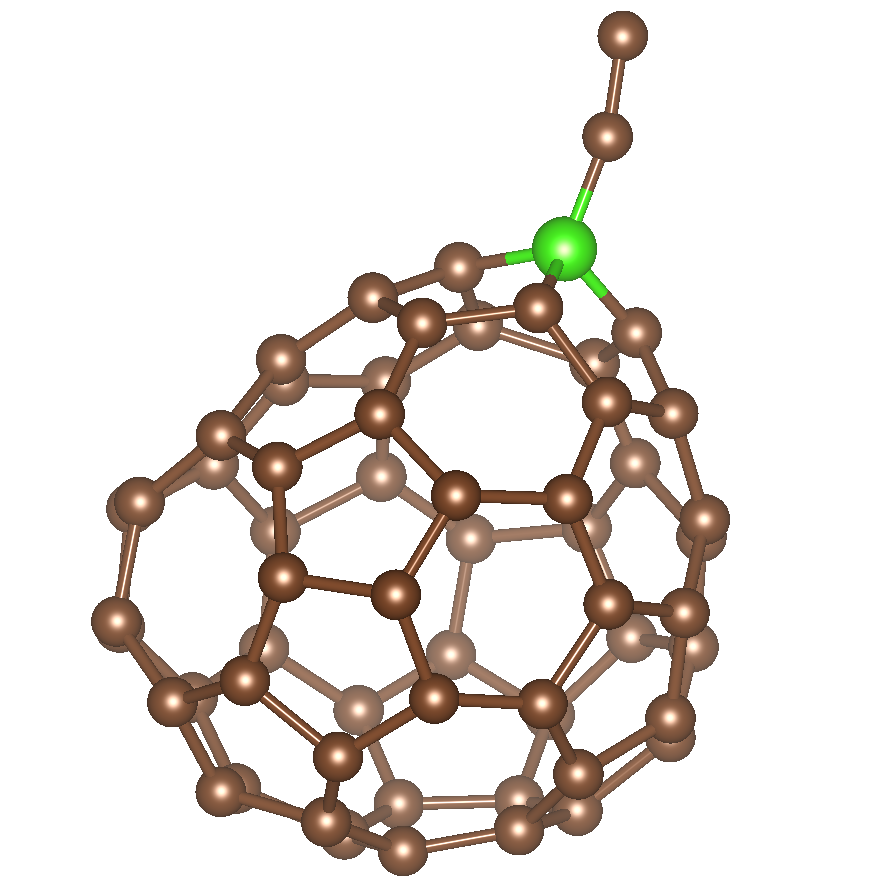}
         \caption{}
         \label{fig:c60corner_016}
     \end{subfigure}
          \begin{subfigure}[b]{0.2\textwidth}
         \centering
         \includegraphics[width=\textwidth]{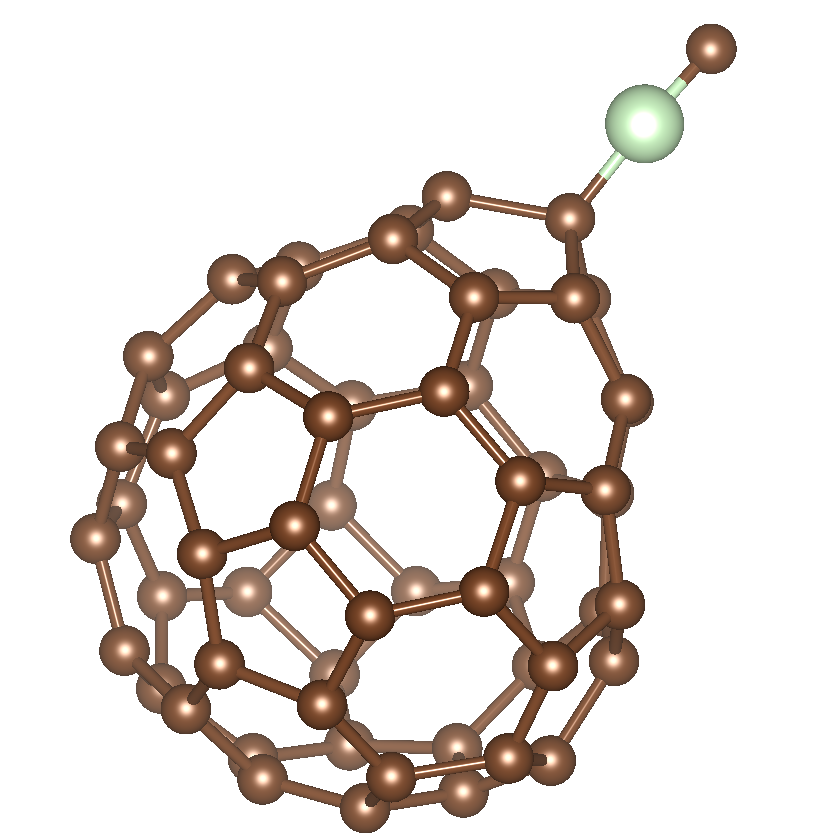}
         \caption{}
         \label{fig:c60corner_017}
     \end{subfigure}
          \begin{subfigure}[b]{0.2\textwidth}
         \centering
         \includegraphics[width=\textwidth]{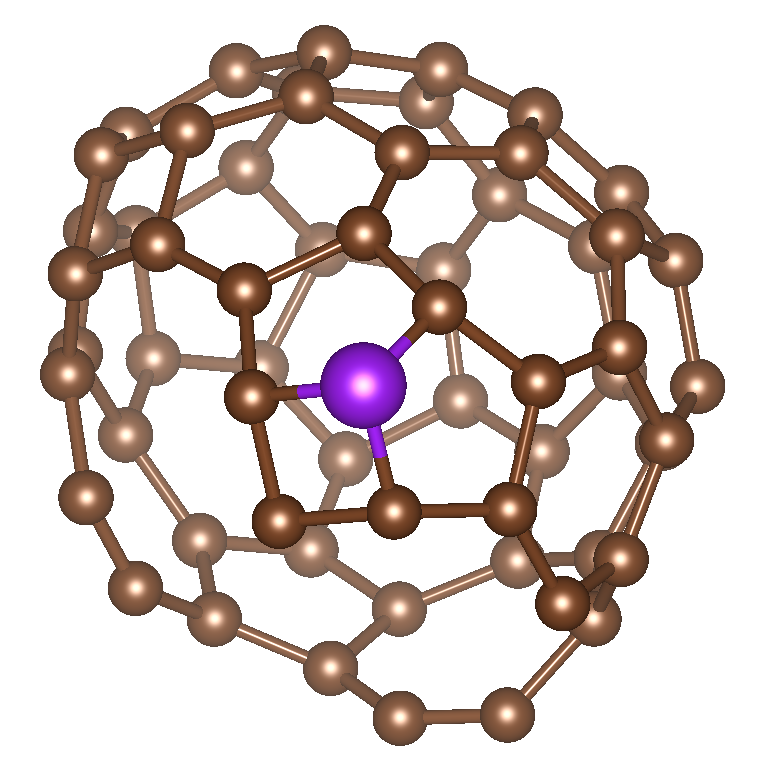}
         \caption{}
         \label{fig:c60corner_018}
     \end{subfigure}
          \begin{subfigure}[b]{0.2\textwidth}
         \centering
         \includegraphics[width=\textwidth]{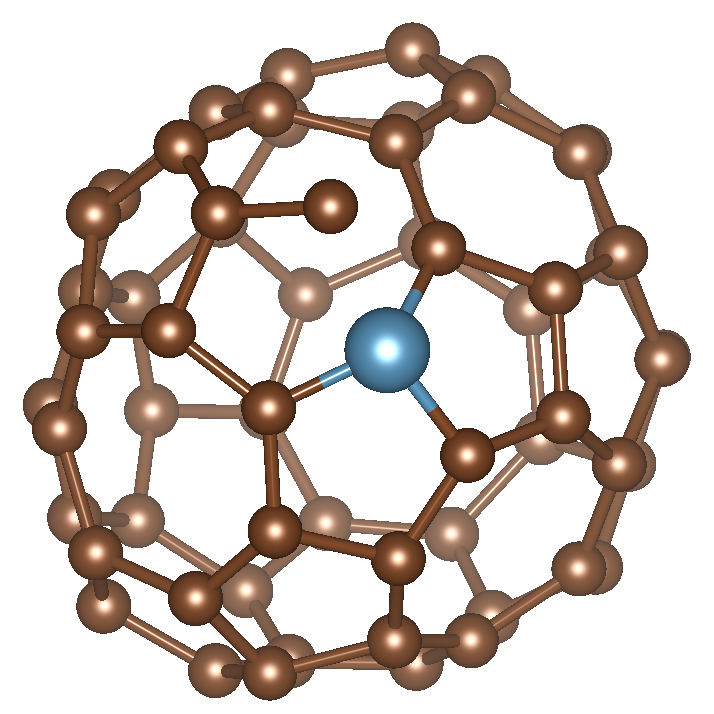}
         \caption{}
         \label{fig:c60corner_019}
     \end{subfigure}
     \caption{The first twenty corners of the LS, i.e. the twenty most distinct atomic environments. The corners are shown in a different color than the rest of the atoms. The relative size and the colors of the atoms is due to the visualization purposes and is not physically important.}
     \label{fig:C60_corners}
\end{figure*}

Our first system to be studied consists of 5000 $C_{60}$ structures, i.e. $5000\times60$ atomic environments, that exhibit several structural motifs including sheets, chains, and cages. These structures were generated by minima hopping \cite{MH} runs coupled to DFTB \cite{aradi2007dftb+}. 
Our aim is to identify the most distinct atomic environments as well as to classify the environments.
 We use OM[sp] with a cutoff radius of $R_c=2w=6$ \AA\ and follow the approach described in section \ref{simplex_method} to generate the LS with $N=60$. 
\\
In Fig.~\ref{fig:C60_corners} we show the first twenty corners of the LS. which represent twenty highly distinct landmark environments in the data set. In agreement with basic chemical intuition the first two corners representing the two most different chemical environments are a four fold 
coordinated atom and a carbon atom at the end of a linear chain with only one nearest neighbor as 
shown in Fig.~\ref{fig:c60corner_001} and Fig.~\ref{fig:c60corner_000}.  
Other two fold coordinated atoms in chains are also represented by higher order corners of the LS as shown in Fig.~\ref{fig:c60corner_005},~\ref{fig:c60corner_016},~\ref{fig:c60corner_017}, and~\ref{fig:c60corner_002}. 
 In Fig.~\ref{fig:c60corner_002} the reference atom is part of a chain but the chain points inside  the cage which shows that our method can distinguish between chains that point inward or outward since it is not based solely on its nearest neighbours, but on its general environment. 


The forth corner of the simplex is an atom with one nearest neighbor and near a hole in the $C_{60}$ shown in Fig.~\ref{fig:c60corner_003}. Other corners of the simplex also clearly represent truly different environments. 
For instance, the 8th corner of the LS shown in Fig.~\ref{fig:c60corner_007} is an atom in a graphite flake and the 16th corner of the LS is an atom in a fragmented part shown in Fig.~\ref{fig:c60corner_015}. Our data set contains only a few fragmented structures in the data set which are of type ~\ref{fig:c60corner_015} and the LS could correctly recognize them as highly distinct environments. \\

\begin{figure}
     \centering
     \includegraphics[width=\columnwidth]{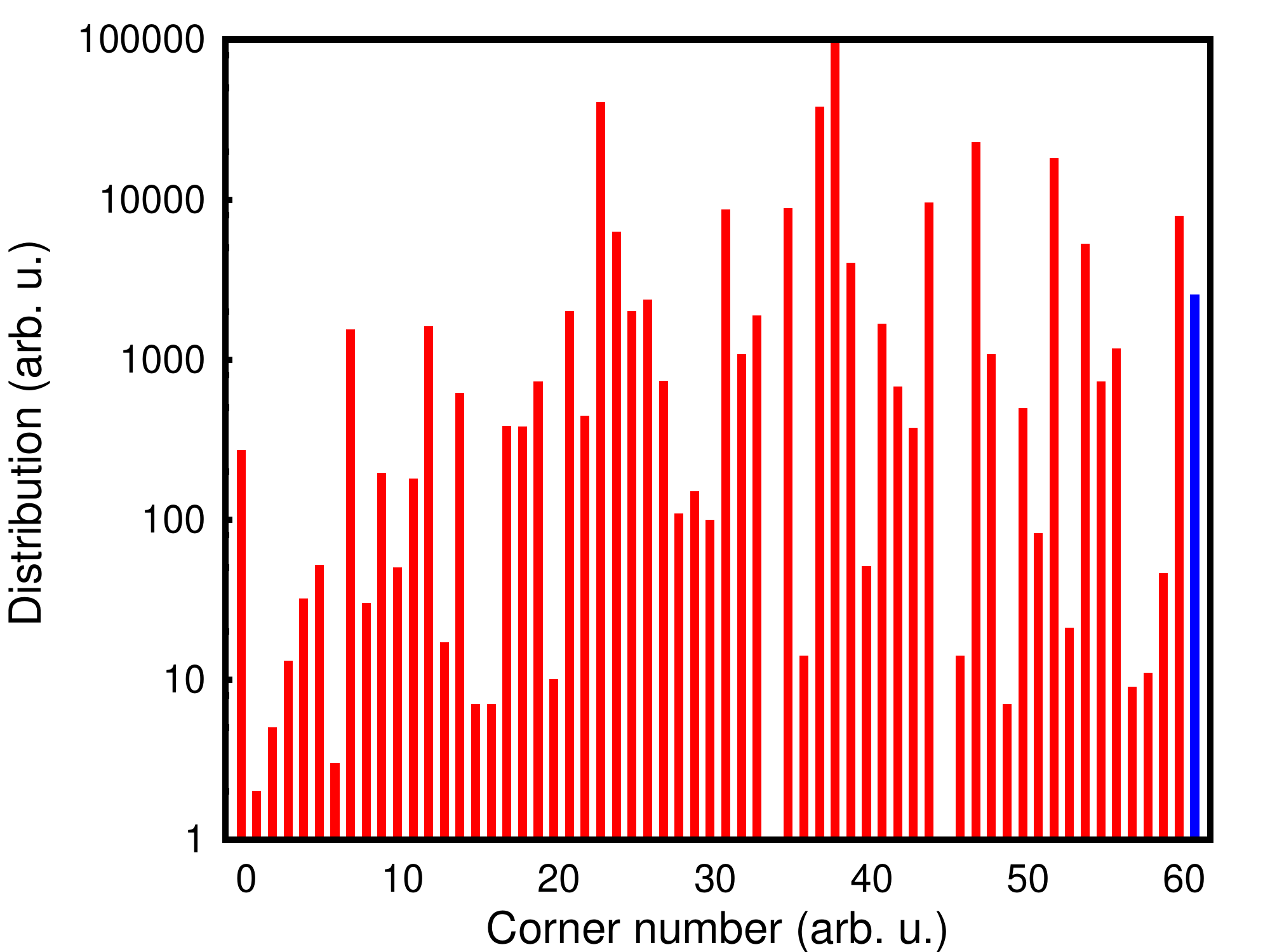}
     \caption{
     The number of atomic environments in the data set of $C_{60}$ structures which are similar to one of the corners of the LS.
     The blue bar represents environments which are not similar to any corner based on the the threshold value $\delta=0.5$.}
     \label{fig:bar}
 \end{figure}

Next, we employ the corners of the LS to analyse structures. Based on the fact that each corner represents highly distinct landmark environments, we can assume that each fingerprint that has a small distance to any of these corners represents an environment that is similar to the corresponding  landmark environment. So, we assign each atomic environment to its closest corner if the fingerprint distance is less than a threshold value $\delta$ which we take to be 0.5. 
With this criterion, we calculate the number of environments which belong to each class as shown in Fig.~\ref{fig:bar}. The environments which do not belong to any corner of the LS, because their  fingerprint distance to the their closest corner is larger than $\delta$, are shown in the blue bar in Fig.~\ref{fig:bar}. Since the first corner is at the origin, Fig.~\ref{fig:bar} starts at zero.
\\
The energetic minimum of the C$_{60}$ molecule is the fullerene molecule. In this structural motif, the atomic environments for all of the carbon atoms are equivalent.
This is not true any more if the fullerene has a so-called Stone-Wales defect~\cite{stone1986theoretical}.
In the following we look at such a structure as well as a 60 atom graphite flake and categorize the atoms according to their fingerprint distance to the landmark environments, i.e. the corners of the LS.
None of the atomic environments of these two structures is actually a landmark environment of the simplex.
For the visualization, we assign a color to each corner of the simplex. All the atomic environments in the data that have a short fingerprint distance to this corner are then shown in this color.

\begin{figure*}[t]
     \centering
     \begin{subfigure}[b]{0.30\textwidth}
         \centering
         \includegraphics[width=\textwidth]{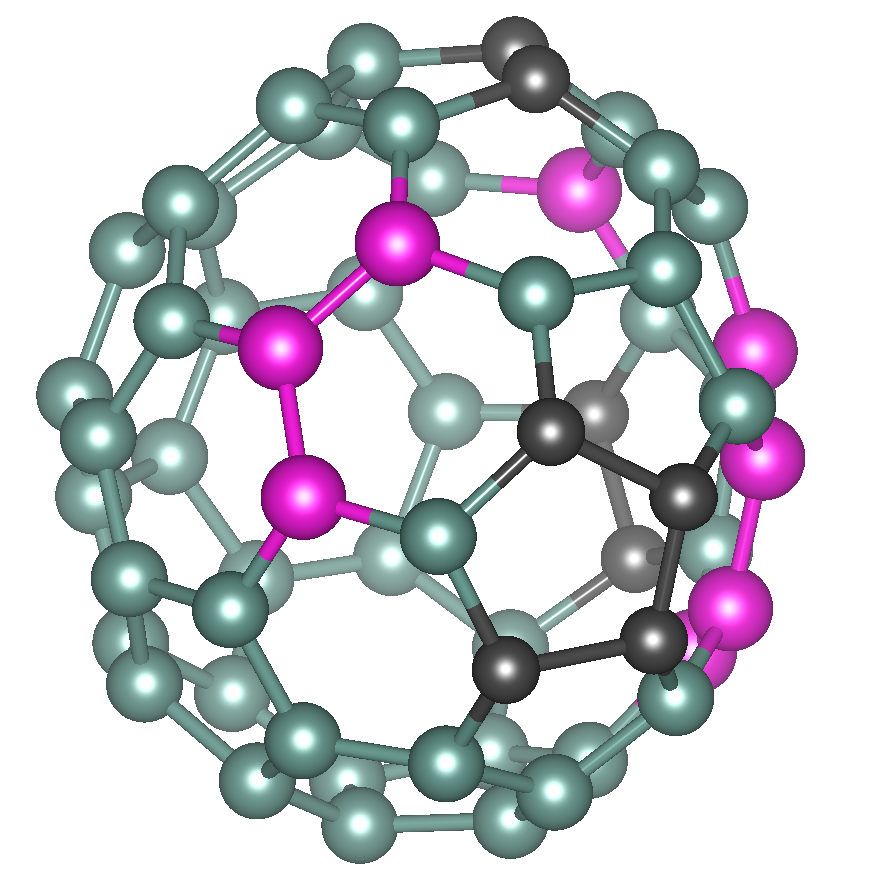}
         \caption{}
         \label{fig:0000}
     \end{subfigure}
     \begin{subfigure}[b]{0.2\textwidth}
         \centering
         \includegraphics[width=\textwidth]{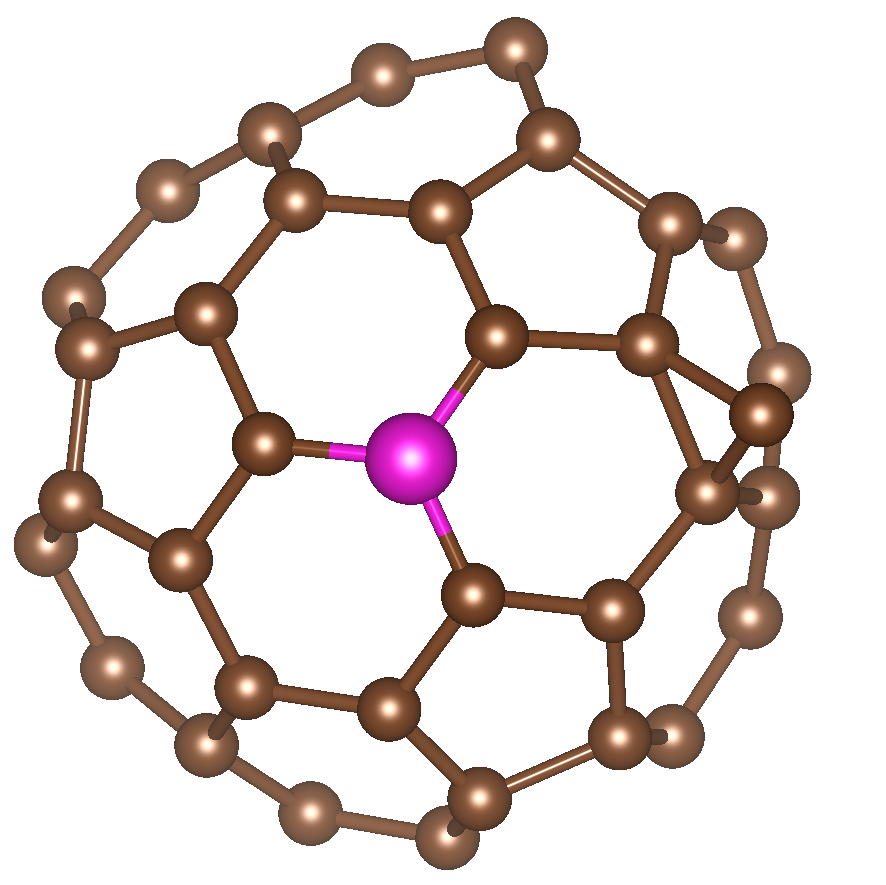}
         \caption{}
         \label{fig:0687}
     \end{subfigure}
     \begin{subfigure}[b]{0.2\textwidth}
         \centering
         \includegraphics[width=\textwidth]{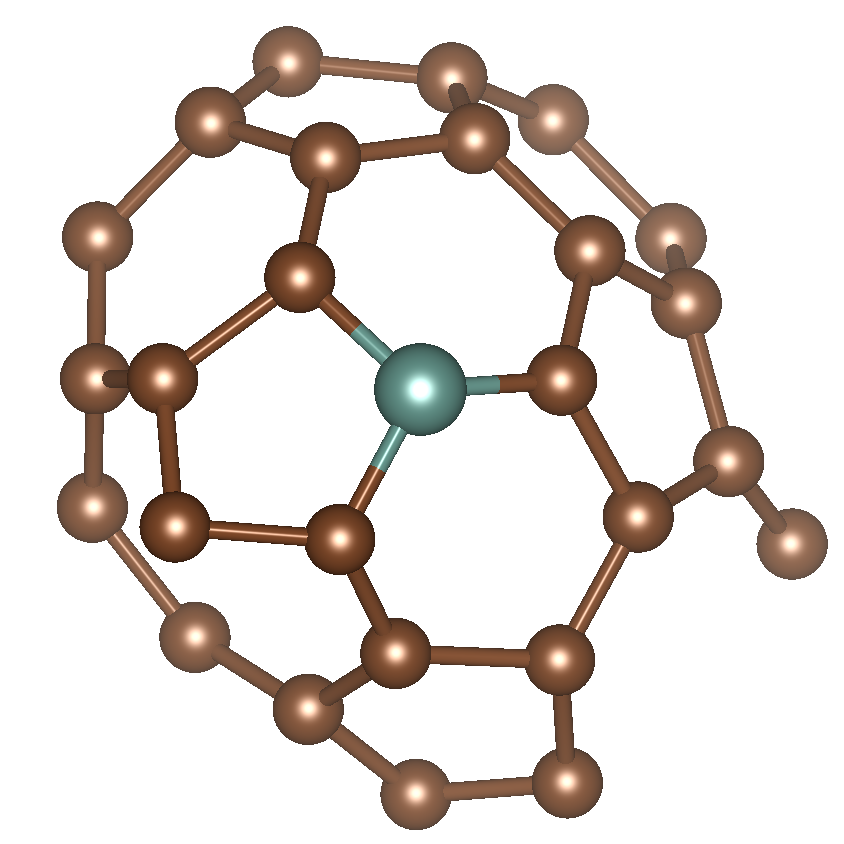}
         \caption{}
         \label{fig:2556all}
     \end{subfigure}
          \begin{subfigure}[b]{0.2\textwidth}
         \centering
         \includegraphics[width=\textwidth]{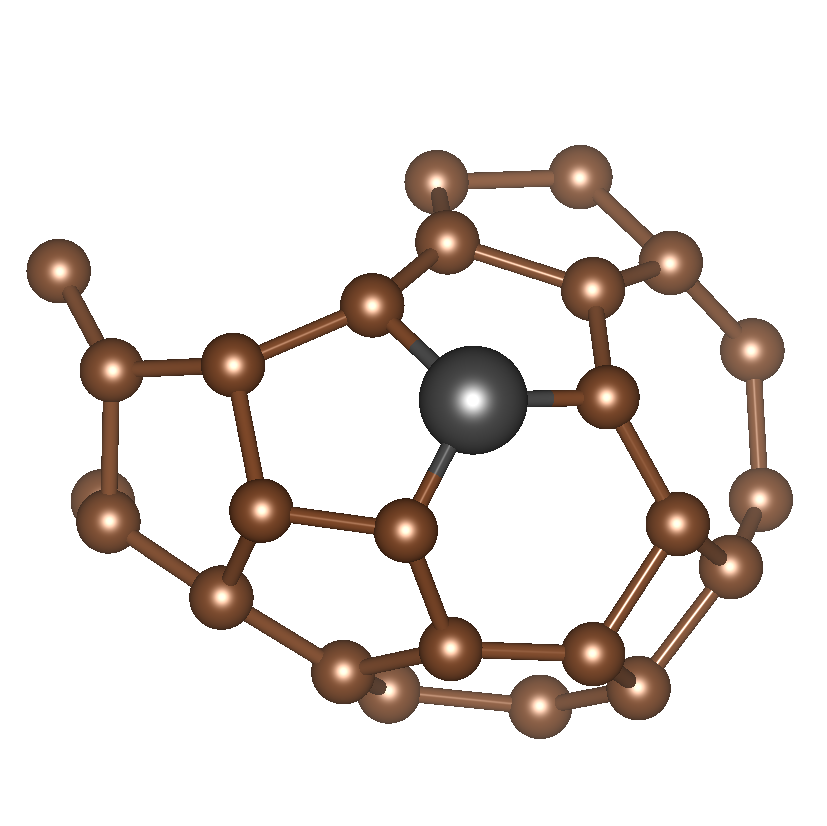}
         \caption{}
         \label{fig:4141all}
     \end{subfigure}
     \caption{
     \textbf{a}) A C$_{60}$ with a Stone-Wales defect: The atoms are colored according to their closest corners which is shown by the same color in the other three images. \textbf{b}) corner 47;  \textbf{c}) corner 38; and \textbf{d}) corner 23 of the LS. 
     }
     
     \label{fig:example1}
\end{figure*}

Our method automatically classifies the atoms of the structure shown in Fig.~\ref{fig:example1} \textbf{a} into three types and we can easily verify  by visual inspection that these three classes are in agreement with chemical intuition:
We see an atom surrounded by two pentagons and one hexagon (corner 47 shown in Fig.~\ref{fig:example1} \textbf{b}); one pentagon and two hexagons (corner 38 shown in Fig.~\ref{fig:example1} \textbf{c}); or three hexagons (corner 23 shown in Fig.~\ref{fig:example1} \textbf{d}).
As can be seen from Fig. \ref{fig:bar}, a large number of atomic environments in our data set are similar to these corners.  

\begin{figure*}[pbth]
     \centering
     \begin{subfigure}[b]{0.30\textwidth}
         \centering
         \includegraphics[width=\textwidth]{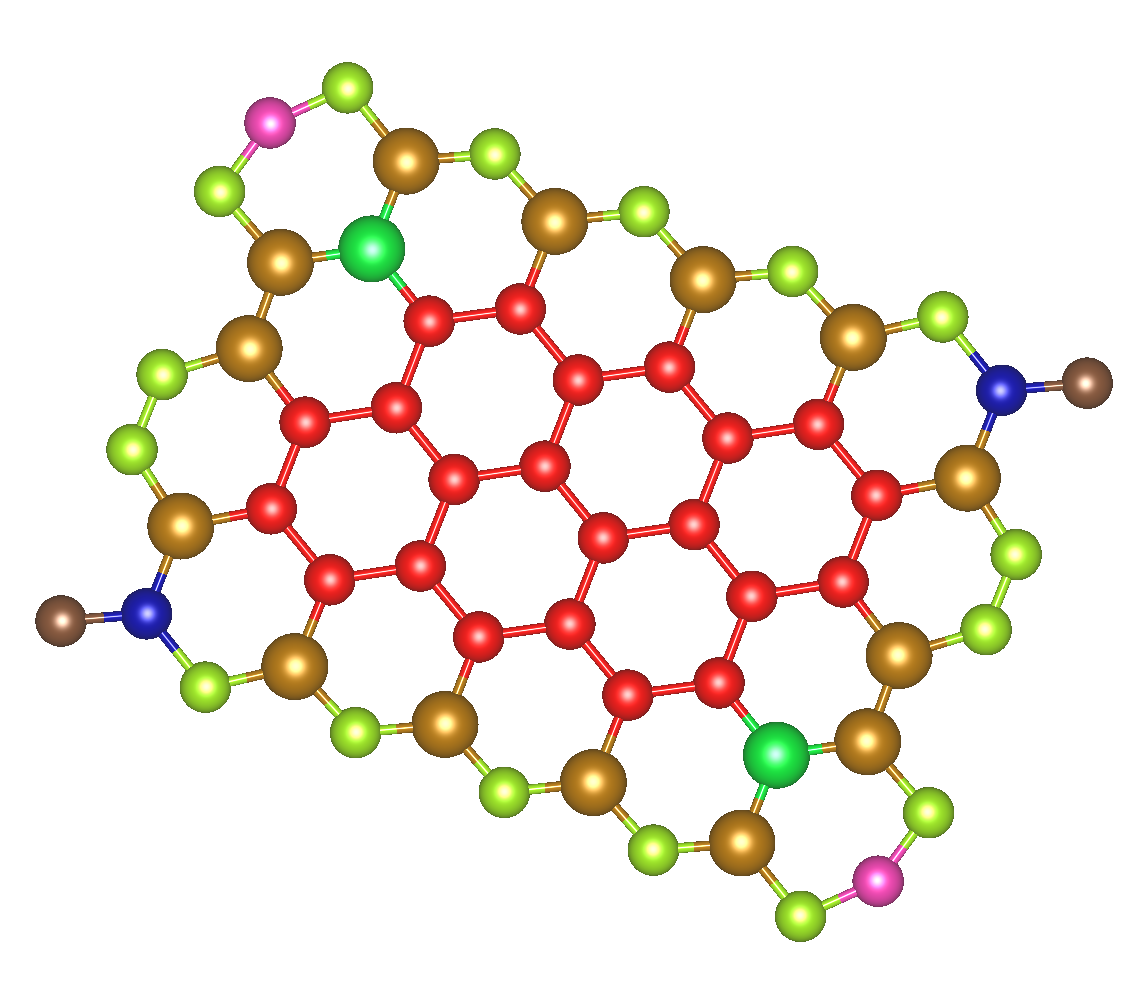}
         \caption{}
         \label{fig:4913}
     \end{subfigure}
     \begin{subfigure}[b]{0.2\textwidth}
         \centering
         \includegraphics[width=\textwidth]{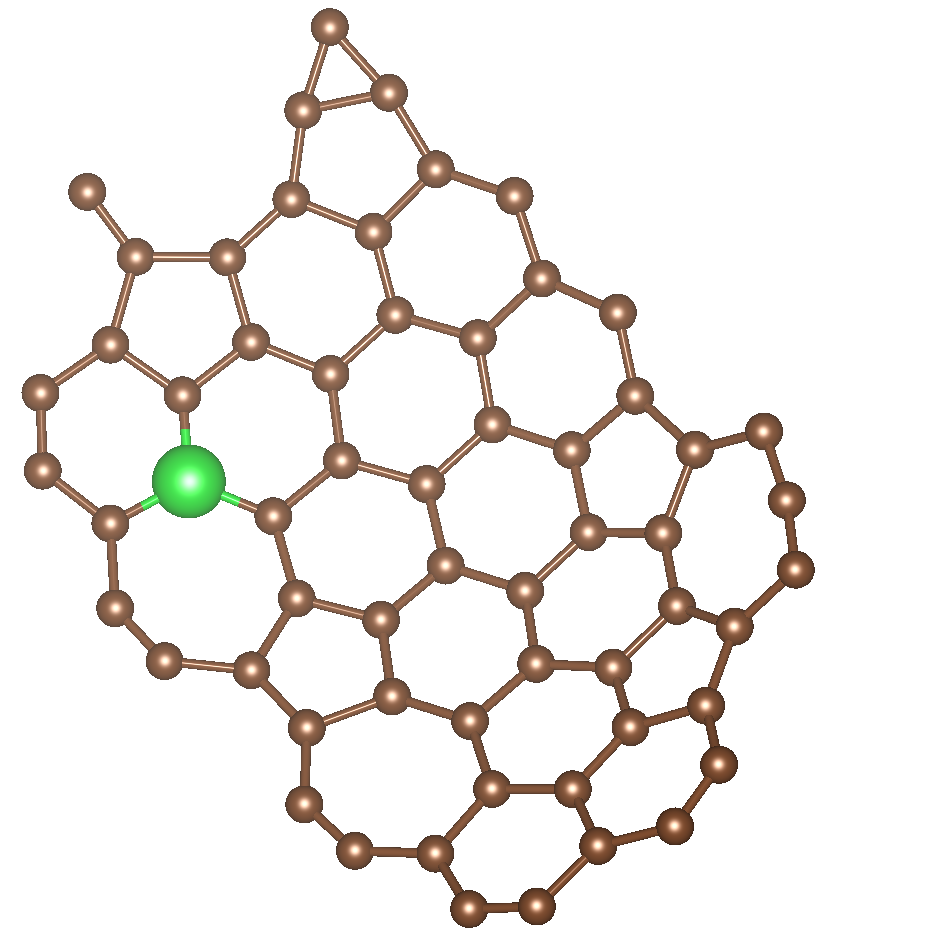}
         \caption{}
         \label{fig:iisim_4737}
     \end{subfigure}
     \begin{subfigure}[b]{0.2\textwidth}
         \centering
         \includegraphics[width=\textwidth]{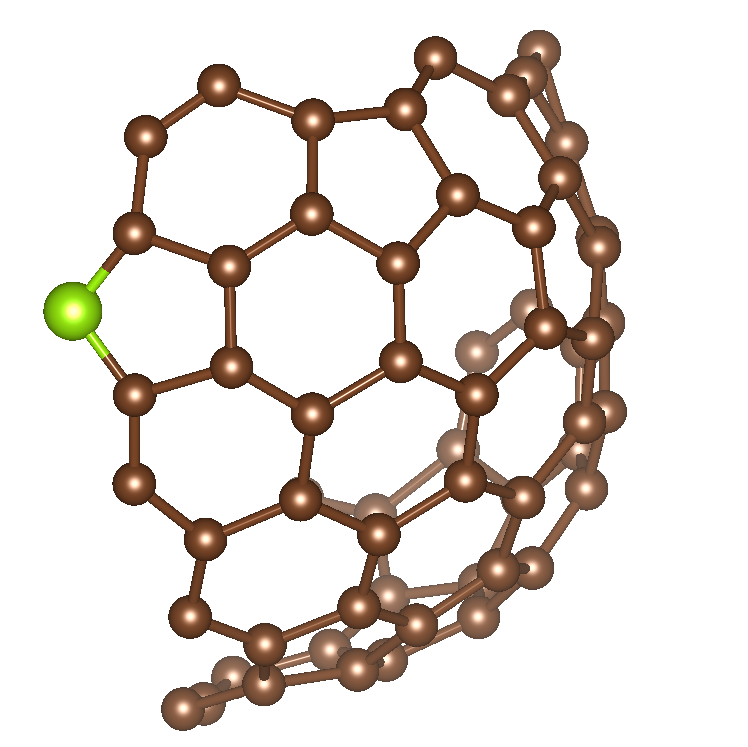}
         \caption{}
         \label{fig:iisim_4721}
     \end{subfigure}
          \begin{subfigure}[b]{0.2\textwidth}
         \centering
         \includegraphics[width=\textwidth]{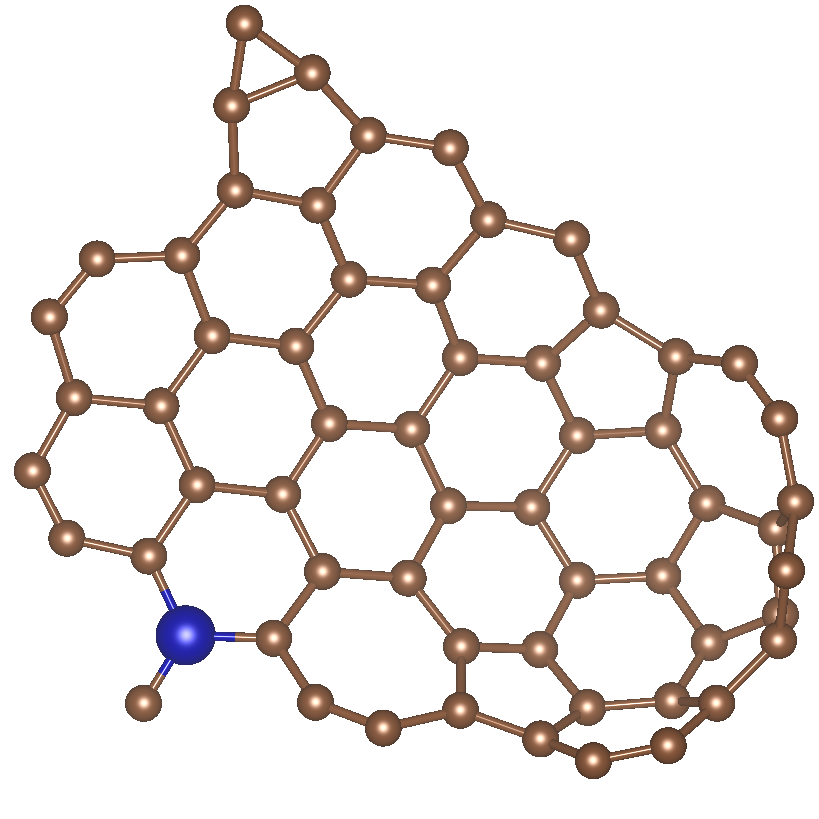}
         \caption{}
         \label{fig:iisim_4810all}
     \end{subfigure}
               \begin{subfigure}[b]{0.2\textwidth}
         \centering
         \includegraphics[width=\textwidth]{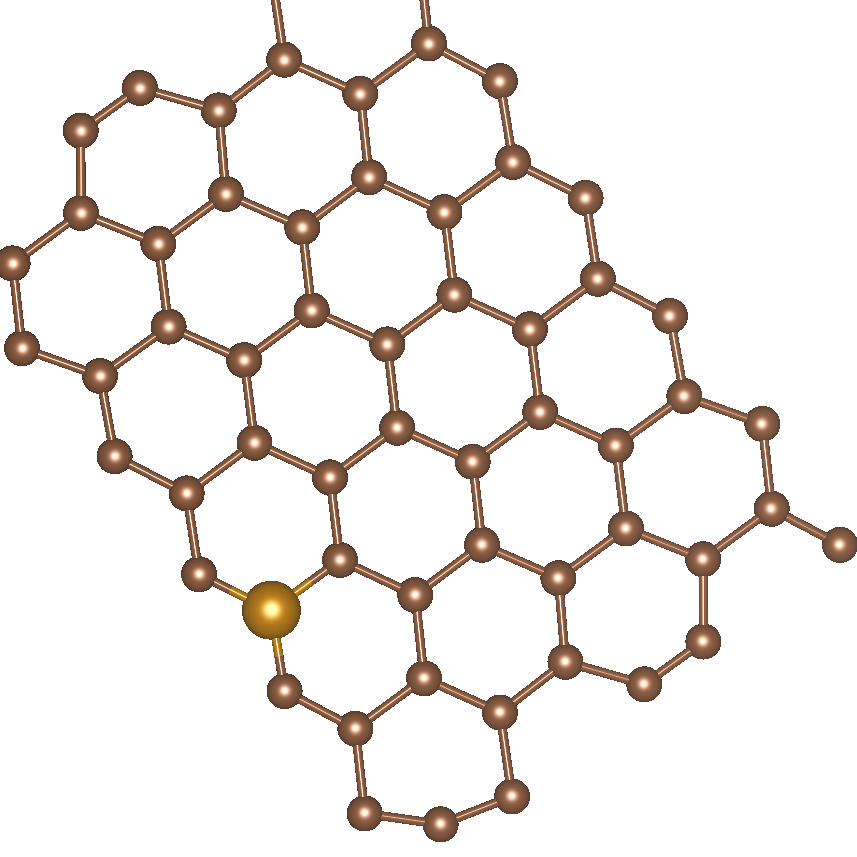}
         \caption{}
         \label{fig:iisim_4914all}
     \end{subfigure}
    \begin{subfigure}[b]{0.2\textwidth}
         \centering
         \includegraphics[width=\textwidth]{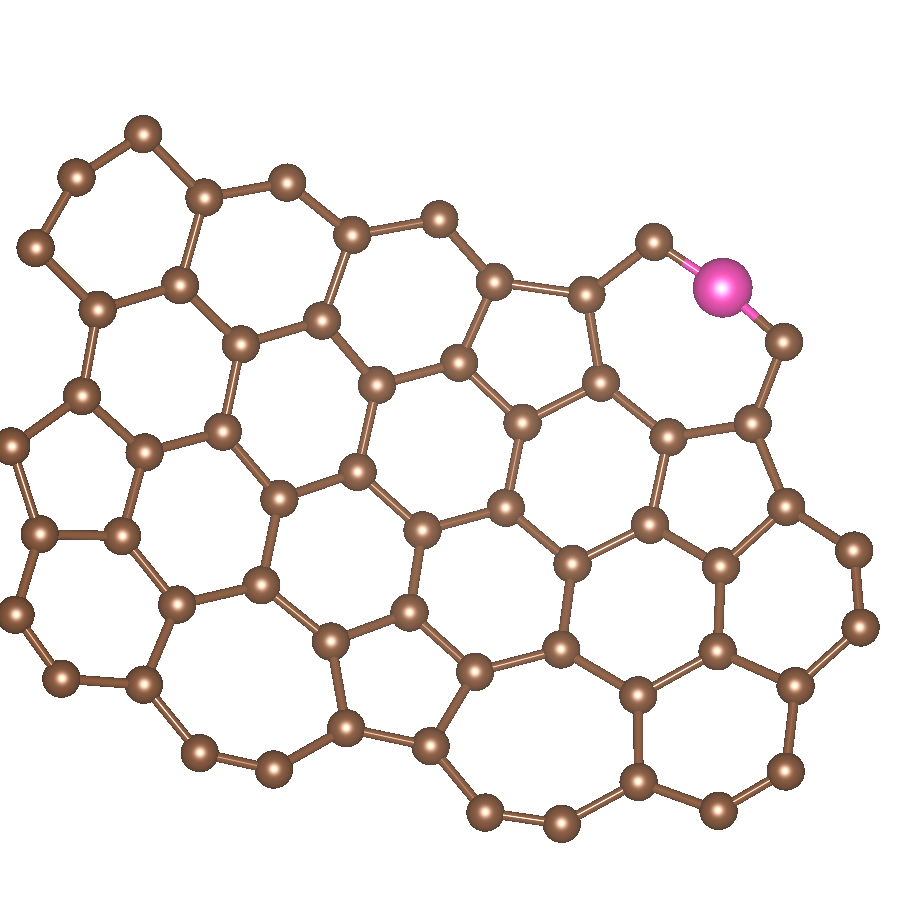}
         \caption{}
         \label{fig:iisim_4705}
     \end{subfigure}
     \begin{subfigure}[b]{0.2\textwidth}
         \centering
         \includegraphics[width=\textwidth]{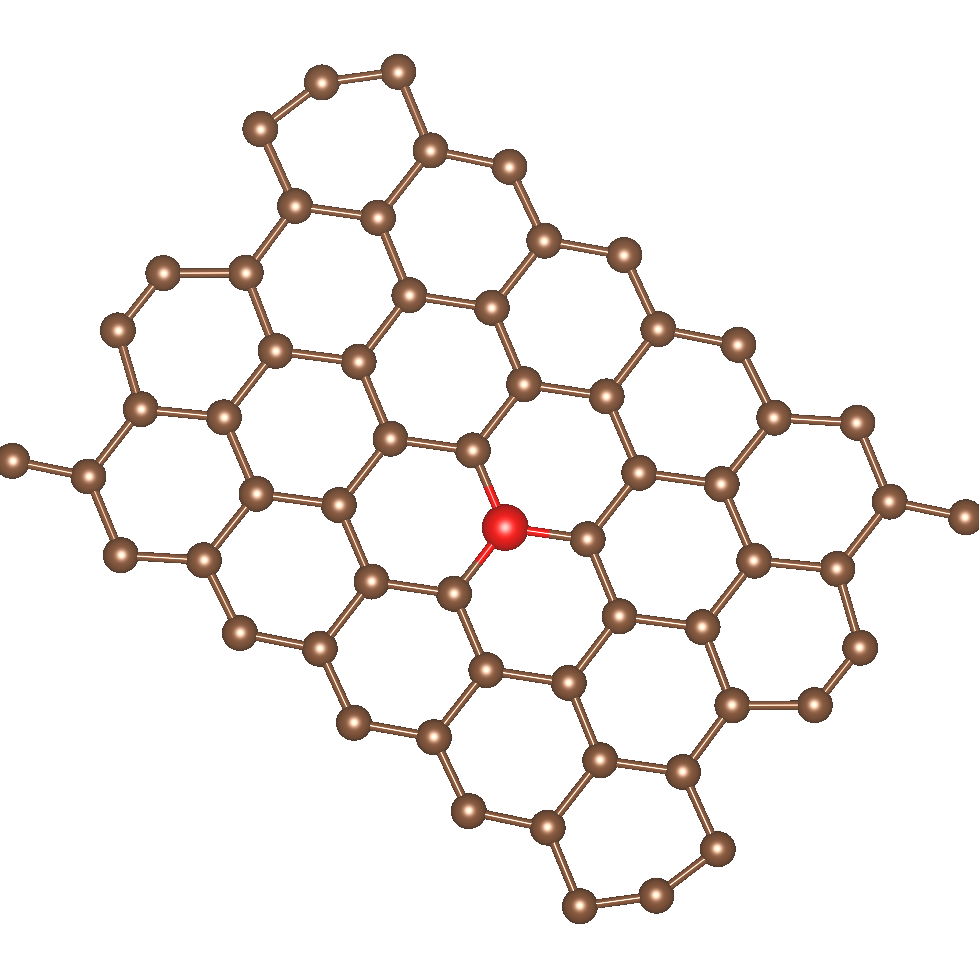}
         \caption{}
         \label{fig:iisim4914all}
     \end{subfigure}
     
     \caption{
     \textbf{a}) A graphite flake whose atoms are colored according to their closest corners. \textbf{b}) corner 55;  \textbf{c}) corner 33;  \textbf{d}) corner 26; \textbf{e}) corner 25; \textbf{f}) corner 9; and \textbf{g}) corner 7 of the LS. 
     }
     
     \label{fig:example2}
\end{figure*}

Another example is shown in Fig.~\ref{fig:example2}. The atoms of the structure in \textbf{a} are similar to one of the 6 different corners of the simplex. These are shown in Fig.~\ref{fig:example2} \textbf{b, c, d, e, f}, and \textbf{g}.
So indeed groups of environments that have a short distances to a landmark environment share similar chemical environments.


\subsection{Grain boundary networks in nanocrystalline Al} \label{sub_Al}

\begin{figure*}[tbph]
     \centering
     \begin{subfigure}[b]{0.30\textwidth}
         \centering
         \includegraphics[width=\textwidth]{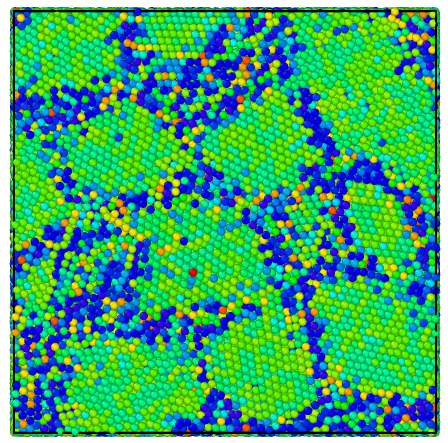}
         \caption{}
         \label{fig:top}
     \end{subfigure}
     \begin{subfigure}[b]{0.3\textwidth}
         \centering
         \includegraphics[width=\textwidth]{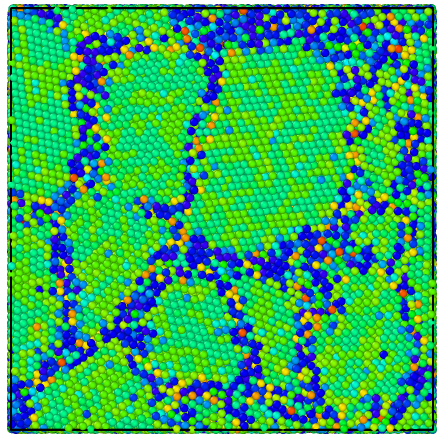}
         \caption{}
         \label{fig:front}
     \end{subfigure}
     \begin{subfigure}[b]{0.3\textwidth}

         \centering
         \includegraphics[width=\textwidth]{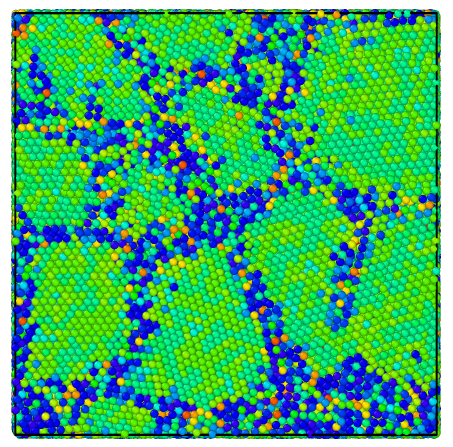}
         \caption{}
         \label{fig:left}
     \end{subfigure}
          \begin{subfigure}[b]{0.3\textwidth}
         \centering
         \includegraphics[width=\textwidth]{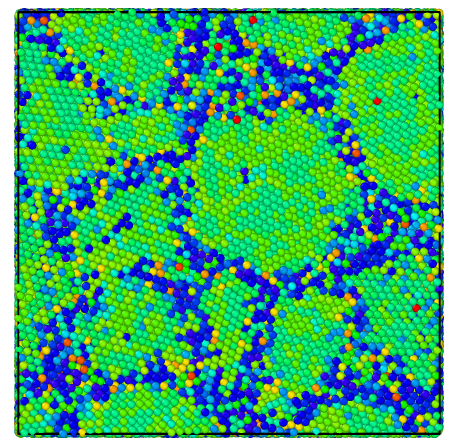}
         \caption{}
         \label{fig:right}
     \end{subfigure}
     \begin{subfigure}[b]{0.3\textwidth}
         \centering
         \includegraphics[width=\textwidth]{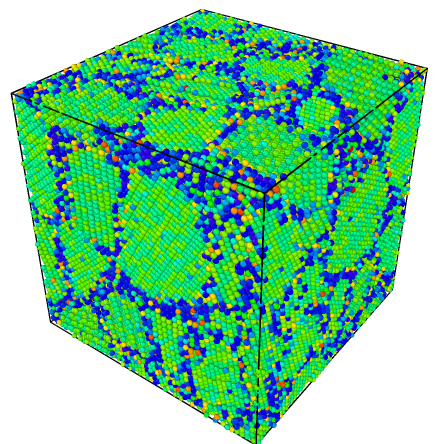}
         \caption{}
         \label{fig:perspective}
     \end{subfigure}
     \begin{subfigure}[b]{0.3\textwidth}
         \centering
         \includegraphics[width=\textwidth]{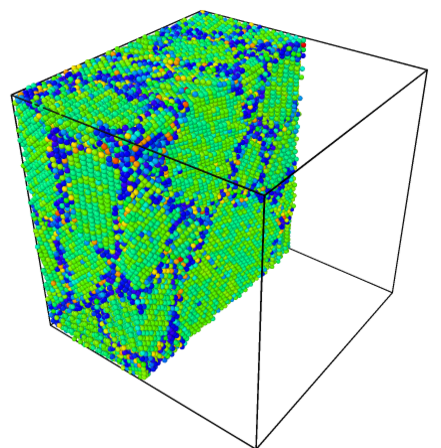}
         \caption{}
         \label{fig:slice}
     \end{subfigure}
     \begin{subfigure}[b]{0.3\textwidth}
         \centering
         \includegraphics[width=\textwidth]{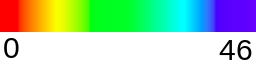}
     \end{subfigure}
     \caption{Nanocrystalline Al containing grain boundaries. The LS is employed to find the grain boundary networks. \ref{fig:top} view from top. \ref{fig:front} view from front. \ref{fig:left} view from left. \ref{fig:right} view from right. \ref{fig:perspective} perspective view. \ref{fig:slice} slice view.  Software Ovito ~\cite{stukowski2009visualization} is used for the visualization.}
     \label{fig:simpl}
\end{figure*}

\begin{figure*}[p]
     \centering
     \begin{subfigure}[b]{0.2\textwidth}
         \centering
         \includegraphics[width=\textwidth]{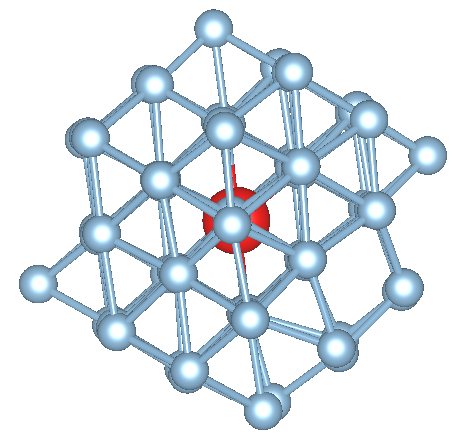}
         \caption{}
         \label{fig:000}
     \end{subfigure}
     \begin{subfigure}[b]{0.2\textwidth}
         \centering
         \includegraphics[width=\textwidth]{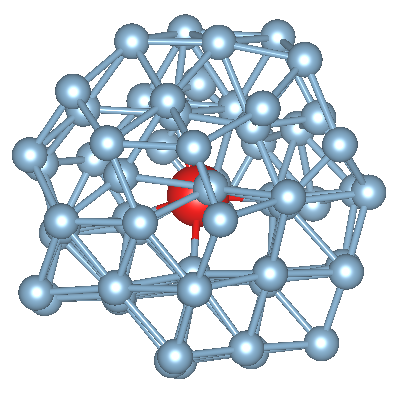}
         \caption{}
         \label{fig:001}
     \end{subfigure}
     \begin{subfigure}[b]{0.2\textwidth}
         \centering
         \includegraphics[width=\textwidth]{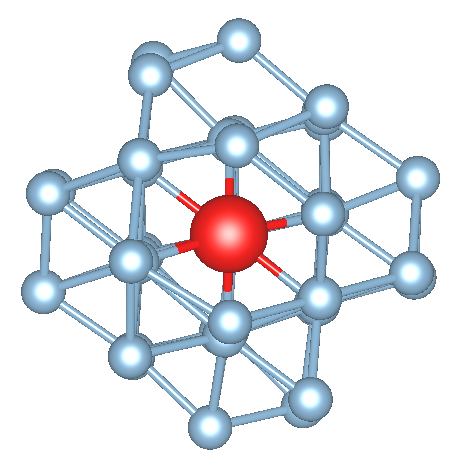}
         \caption{}
         \label{fig:002}
     \end{subfigure}
     \begin{subfigure}[b]{0.2\textwidth}
         \centering
         \includegraphics[width=\textwidth]{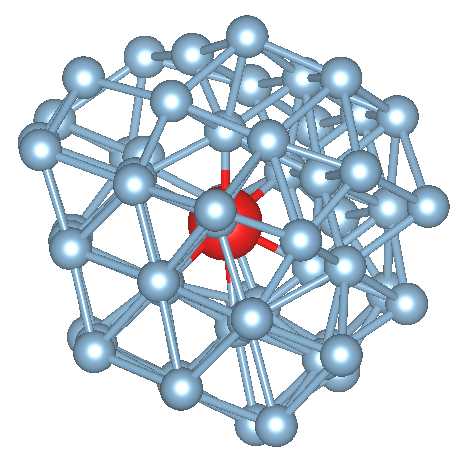}
         \caption{}
         \label{fig:003}
     \end{subfigure}
     \begin{subfigure}[b]{0.2\textwidth}
         \centering
         \includegraphics[width=\textwidth]{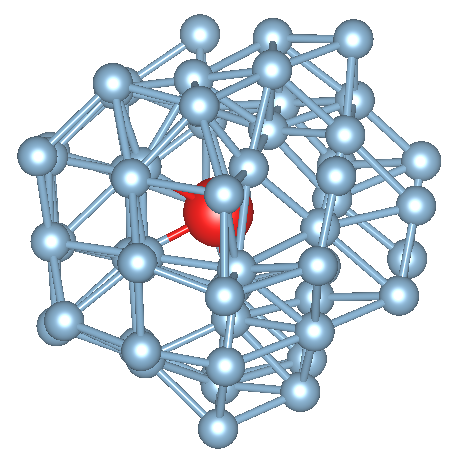}
         \caption{}
         \label{fig:004}
     \end{subfigure}
     \begin{subfigure}[b]{0.2\textwidth}
         \centering
         \includegraphics[width=\textwidth]{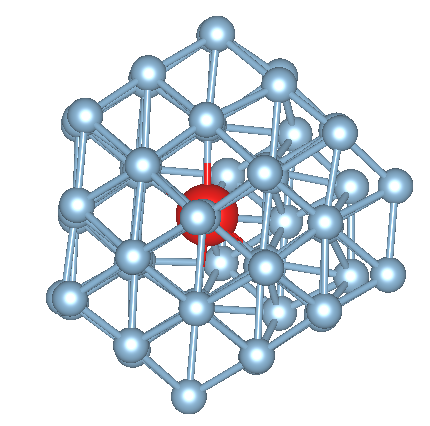}
         \caption{}
         \label{fig:005}
     \end{subfigure}
      \begin{subfigure}[b]{0.2\textwidth}
         \centering
         \includegraphics[width=\textwidth]{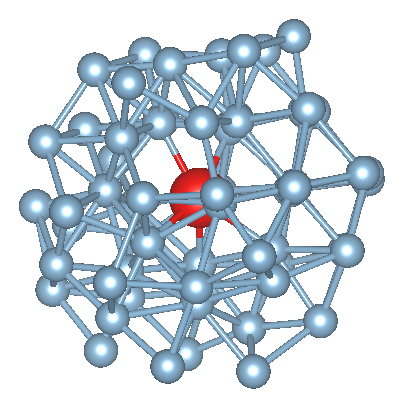}
         \caption{}
         \label{fig:006}
     \end{subfigure}
      \begin{subfigure}[b]{0.2\textwidth}
         \centering
         \includegraphics[width=\textwidth]{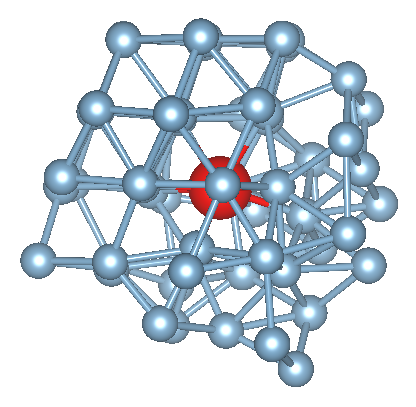}
         \caption{}
         \label{fig:007}
     \end{subfigure}
      \begin{subfigure}[b]{0.2\textwidth}
         \centering
         \includegraphics[width=\textwidth]{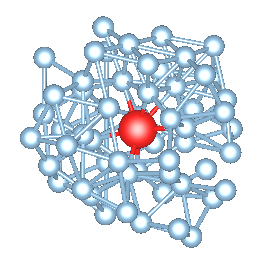}
         \caption{}
         \label{fig:008}
     \end{subfigure}
      \begin{subfigure}[b]{0.2\textwidth}
         \centering
         \includegraphics[width=\textwidth]{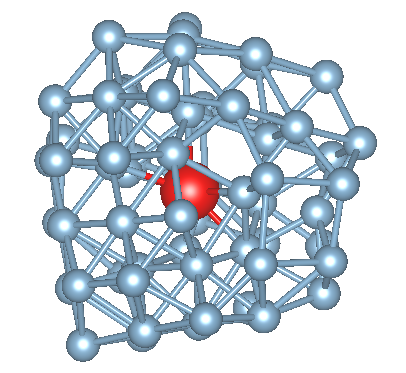}
         \caption{}
         \label{fig:009}
     \end{subfigure}
      \begin{subfigure}[b]{0.2\textwidth}
         \centering
         \includegraphics[width=\textwidth]{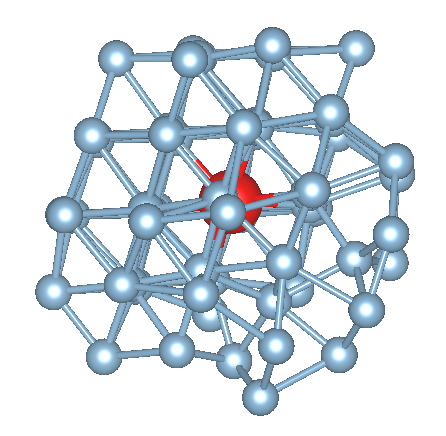}
         \caption{}
         \label{fig:010}
     \end{subfigure}
      \begin{subfigure}[b]{0.2\textwidth}
         \centering
         \includegraphics[width=\textwidth]{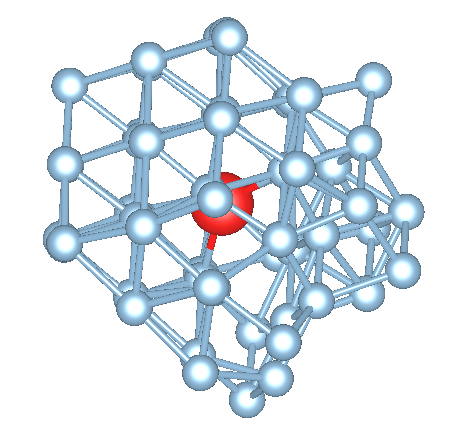}
         \caption{}
         \label{fig:011}
     \end{subfigure}
      \begin{subfigure}[b]{0.2\textwidth}
         \centering
         \includegraphics[width=\textwidth]{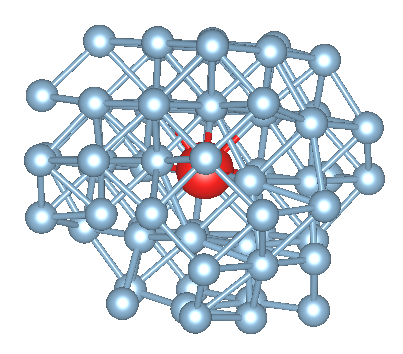}
         \caption{}
         \label{fig:012}
     \end{subfigure}
     \begin{subfigure}[b]{0.2\textwidth}
         \centering
         \includegraphics[width=\textwidth]{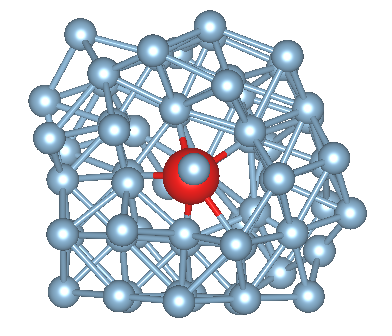}
         \caption{}
         \label{fig:013}
     \end{subfigure}
     \begin{subfigure}[b]{0.2\textwidth}
         \centering
         \includegraphics[width=\textwidth]{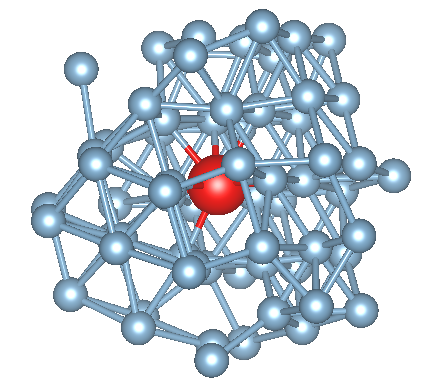}
         \caption{}
         \label{fig:014}
     \end{subfigure}
     \begin{subfigure}[b]{0.2\textwidth}
         \centering
         \includegraphics[width=\textwidth]{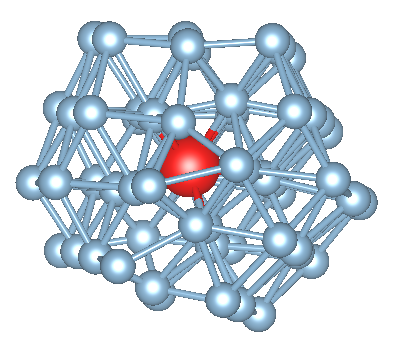}
         \caption{}
         \label{fig:015}
     \end{subfigure}
     \begin{subfigure}[b]{0.2\textwidth}
         \centering
         \includegraphics[width=\textwidth]{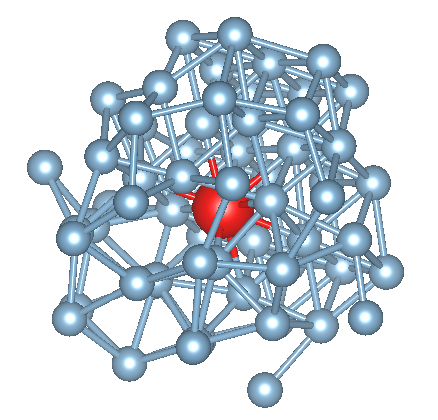}
         \caption{}
         \label{fig:016}
     \end{subfigure}
          \begin{subfigure}[b]{0.2\textwidth}
         \centering
         \includegraphics[width=\textwidth]{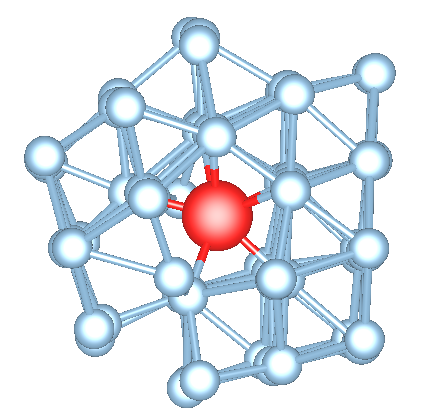}
         \caption{}
         \label{fig:017}
     \end{subfigure}
          \begin{subfigure}[b]{0.2\textwidth}
         \centering
         \includegraphics[width=\textwidth]{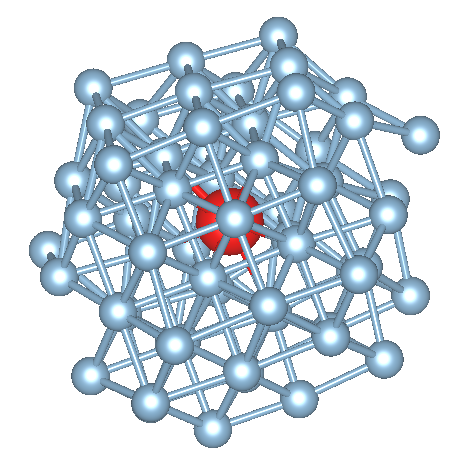}
         \caption{}
         \label{fig:018}
     \end{subfigure}
          \begin{subfigure}[b]{0.2\textwidth}
         \centering
         \includegraphics[width=\textwidth]{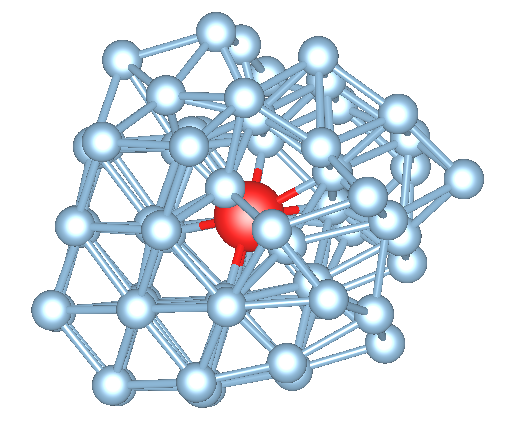}
         \caption{}
         \label{fig:019}
     \end{subfigure}
     \caption{The first twenty most distinctive atomic environments in the nanocrystalline Al found by the LS. Red atoms are the central atoms whose local environment is one of the corners of the simplex and the atoms in their vicinity are depicted in blue. 
     }
     \label{fig:corners}
\end{figure*}

In our second application, we study a nanocrystalline Al aggregate with 255064 atoms containing grain boundary networks. The details on the generation of the nanocrystalline Al used here can be found elsewhere~\cite{piaggi2017entropy}. 
We use the OM[s] fingerprint with a cutoff radius of $R_c=5$ \AA\ to build the LS. We take $N=46$ which is the same as the length of the fingerprint.
Having generated the LS, we assign 
a different color to each of the corners of the simplex for the following visualizations.
These corners are the most distinct environments in the nanocrystalline Al, i.e. each corner can represent a class of diverse environments in the data. 
We again categorize the atoms in the system according to their similarity to the corners of the LS and assign them the same color as the corners they resemble most.
Visual inspection of Fig. \ref{fig:simpl} shows that the simplex method can find all the grain boundary networks, in agreement with the findings of Piaggi~\cite{piaggi2017entropy}. In addition, it can also recognize differences between different grain boundaries and find different kinds of ordered-disordered phases as shown in Fig.~\ref{fig:corners}. 
 
In Fig.~\ref{fig:corners} we showed the first 20 corners of the LS. Fig. \ref{fig:000} shows a perfect crystalline FCC phase. Figs. \ref{fig:002} and \ref{fig:017} show the defective crystalline FCC phases where one nearest neighbor of the central atom is missing. The corners shown in Figs. \ref{fig:004}, \ref{fig:013}, \ref{fig:015}, and \ref{fig:018} correspond to atoms on a twisted grain  boundary. The configurations from Figs. \ref{fig:001}, \ref{fig:003}, \ref{fig:007}, \ref{fig:011}, and \ref{fig:019} represent environments located on the boundary between ordered and disordered phases. Finally, some corners of the LS  
represent atoms in disordered phases such as those shown in Figs. \ref{fig:008} and \ref{fig:009}. 


\subsection{The compression of the fingerprints} \label{sec:discussion}

\begin{figure*}
    \centering
    \includegraphics[width=\textwidth]{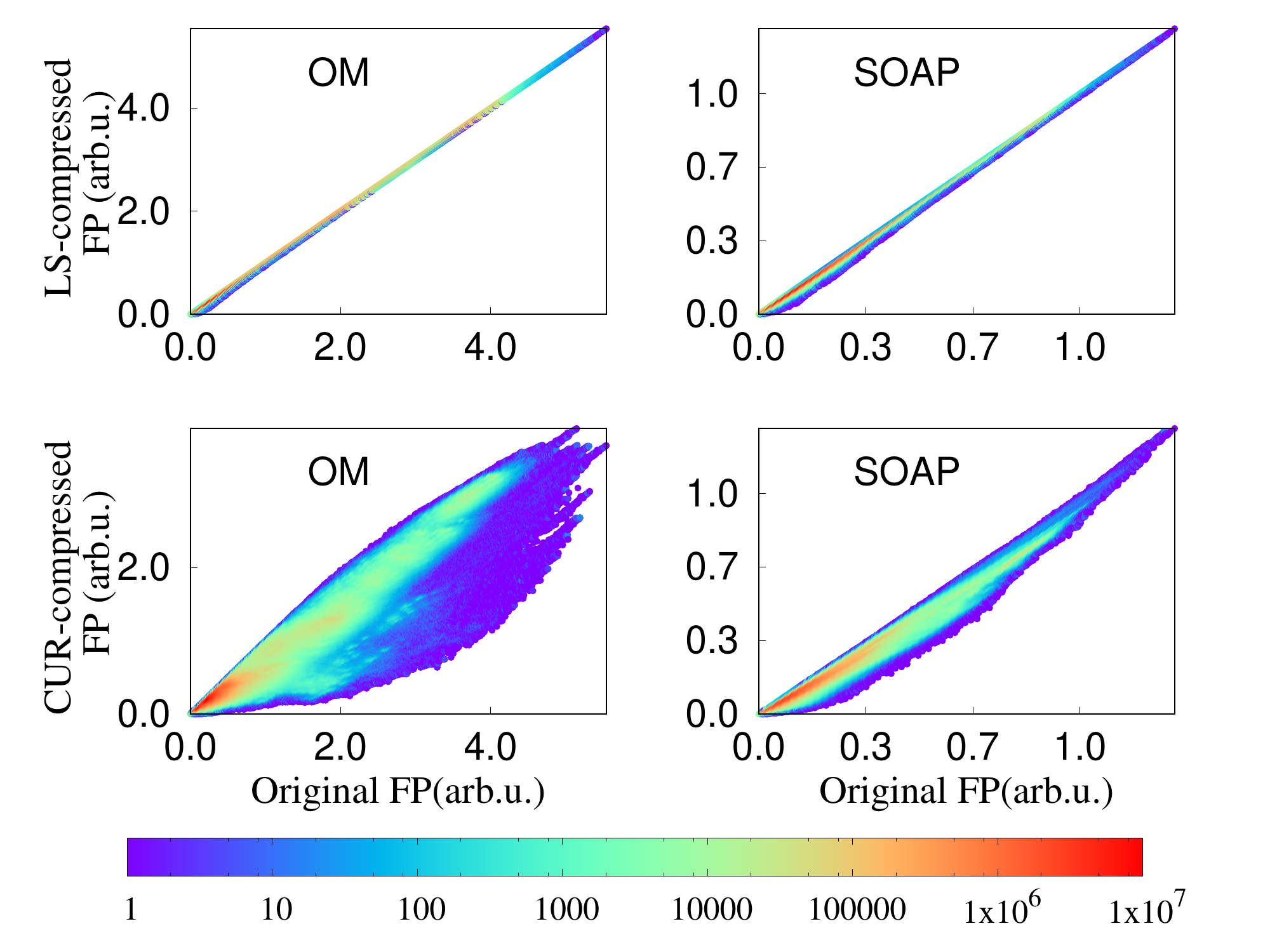}
    \caption{The correlation between the original fingerprints and CUR and LS-reduced fingerprints for OM and SOAP. The length of the reduced fingerprints is $l=16$ while the length of the original fingerprint $L$ is 240 for OM and 325 for SOAP.}
    \label{fig:corr-plot}
\end{figure*}


In section~\ref{simplex_method} we showed that once the LS is found, the original fingerprints can be projected onto the LS. In this section we will show that these projections can be regarded as a new fingerprint whose length is much shorter than the original fingerprint while containing most of the information of the original fingerprint. This is an example of data compression, a problem  for which many algorithms are available such as CUR \cite{mahoney2009cur} decomposition. Assuming that $F$ is the fingerprint matrix with dimension $L\times N'$ where $L$ is the length of the fingerprint and $N'$ is the number of atomic environments $N'=N_{env}$, i.e. $i$th column of $F$ contains the fingerprint vector of atomic environment $i$, one can write $F\sim CUR$ in which $C$ and $R$ contain $k$ selected columns and rows of $F$ and $U=C^+FR^+$ where $A^+$ indicates the pseudo-inverse of $A$ and $k < r=rank(F)$.
In order to find the reduced selected number of rows of matrix $F$, one writes its SVD decomposition as $F=\bar {U}DV^T$, where $\bar{U}$ (left singular matrix) and $V$ (right singular matrix) are $L\times L$ and $N'\times N'$ unitary matrices and $D$ is a $L\times N'$ rectangular diagonal matrix with non-negative real numbers on the diagonal. The diagonal entries of $D$ are known as the singular values of $F$. Then the leverage score for each row $i$ is calculated as $\pi_i=\frac{1}{k}\sum_{\xi=1}^{k}{(u^\xi_i)^2}$ where $u^\xi_i$ is the $i$th component of $\xi$th left singular vector and $k$ is the number of rows that should be selected. Frequently, rows are selected with probability proportional to the leverage score. We employed a deterministic method \cite{imbalzano2018automatic,ceriotti2020machine} and select the row with the highest leverage score at each time. Then, the selected row is removed from the matrix and the rest of the rows become orthogonalized with respect to it. To select other rows this procedure is repeated. The selected rows are the most important features. One can also select columns of the matrix $F$, i.e. the most important atomic environments by following the same procedure but for $F^T$. The selected rows and column are stored in $R$ and $C$ respectively.  \\
In the following, we employ the LS and CUR method to reduce the length of the fingerprint by selecting the components  of the fingerprint that contain the most important information.

In order to investigate whether the compressed fingerprint conserves the information encoded in the original fingerprint, we correlate all the pairwise fingerprint distances obtained by the original and compressed fingerprints \cite{parsaeifard2020assessment}. 

Obviously fingerprint distances that are large with the original fingerprint should remain large  with 
the compressed fingerprint. In the same way short distances should remain short.
If this is the case all the points in a correlation plot between the fingerprint distances arising from  the original and the compressed fingerprint will lie on or close to the diagonal. 
 If there are points far away from the diagonal and in particular if some fingerprint distances 
 of the compressed fingerprint are small whereas the original distances are large, there is a loss of information.
 

In Fig.~\ref{fig:corr-plot} we show the correlation plot between the original fingerprints and the CUR-reduced and LS-reduced fingerprints using OM and SOAP~\cite{bartok2013representing} for 
our above-mentioned test of 1000  C$_{60}$ clusters with  $1000\times 60$ atomic environments.
We used the same fingerprint parameters for OM as in section~\ref{sub_C60}. For SOAP, we used the following parameters: $l_{max}=n_{max}=8$, $r_\delta=4.0$ \AA, $\sigma=0.5$ \AA. The cutoff radius is the same $6$ \AA\ in both OM and SOAP.
The software QUIP~\cite{quippy} is used to generate the SOAP fingerprints. The length $L$ of the original fingerprints is 240 for OM and 325 for SOAP. We reduced the length of the fingerprints to $l=16$. As can be seen in Fig.~\ref{fig:corr-plot}, the correlation is perfectly diagonal in the case of LS which indicates that vast majority of the information of the original fingerprint is retained in the LS-reduced fingerprint. There are however some deviations from the diagonal in the correlation plot between the original fingerprint and CUR-reduced fingerprint which indicates that some information is lost in the CUR-decomposition.

\section{conclusion}
We have introduced an algorithm to construct a largest volume simplex in the space spanned by a large set of atomic environment fingerprint vectors. 
The number of corners of this simplex gives the effective dimension of the fingerprint vector space.
The corners themselves represent landmark environments that can be used to analyse structures with a large number of atoms in a fully 
automatic way. So, in contrast to other methods, it is not necessary to include into our analysis tool criteria that are based on human expectations of what kind of environments are expected to be 
encountered in this system. We show that this analysis method can be used to detect grain boundaries and other typical environments in multi-grain metallic systems and to classify atomic environments  
in a carbon cluster in a way that is consistent with basic chemical intuition. 
Since only those components of the fingerprint vector that are inside the space spanned by the LS are relevant, projecting the fingerprint into the space spanned by the simplex reduces the length of the fingerprint without any significant loss of information. Therefore the method can also be used as a data compression method for fingerprints.

\section{acknowledgments} 
We thank Dr. Pablo Piaggi for providing us with the nanocrystalline Al data. The authors acknowledge that this research was supported by NCCR MARVEL and funded by the Swiss National Science Foundation. Structures were visualized using VESTA~\cite{momma2011vesta} and Ovito~\cite{stukowski2009visualization} packages. The calculations were performed on the computational resources of the Swiss National Supercomputer (CSCS) under project s963 and on the Scicore computing center of the University of Basel. 
\bibliography{main}

\end{document}